\documentclass[conference]{IEEEtran}

\newif\ifanonymous

\usepackage[
 colorlinks=true,
 allcolors=black,
]{hyperref}
\usepackage[all]{hypcap} %

\usepackage[numbers,sort]{natbib}
\usepackage{flushend} %
\usepackage{graphicx}
\usepackage{xcolor}
\usepackage{amsmath}
\usepackage{mathtools} %
\usepackage{inconsolata}

\usepackage{booktabs}

\usepackage{shellesc}

\usepackage[colorinlistoftodos,prependcaption,textsize=small]{todonotes}

\usepackage{tabularx,multirow} %
\usepackage[english]{babel}
\usepackage[autostyle]{csquotes} 
\usepackage{color, colortbl}
\usepackage{caption}
\usepackage{subcaption}
\usepackage{placeins}
\usepackage{ulem}
\usepackage{bm}
\usepackage[super]{nth}
\usepackage{enumitem} %
\setlist[enumerate]{topsep=3pt,itemsep=0ex}
\setlist[itemize]{topsep=3pt,itemsep=0ex}

\usepackage{tikz}
\usetikzlibrary{external}
\usetikzlibrary{positioning,patterns}
\usetikzlibrary{shapes,arrows}
\usetikzlibrary{calc}
\newcommand*\circled[1]{\tikz[baseline=(char.base)]{
		\node[shape=circle,draw,inner sep=0.5pt,font=\footnotesize\sffamily] (char) {#1};}}
	\newcommand{\bluebox}{\textcolor{blue}{\rule{.5cm}{.2cm}}}

\usepackage{pgfplotstable}
\usepackage{pgfplots}
\usepgfplotslibrary{groupplots}
\pgfplotsset{compat=1.17}
\usepgfplotslibrary{patchplots}
\pgfplotstableset{col sep=space}

\usepackage{arydshln}
\makeatletter
\renewcommand{\todo}[2][]{\tikzexternaldisable\@todo[#1]{#2}\tikzexternalenable}
\makeatother

\def\Tool{XFL}
\def\Embeddings{\textsc{Dexter}}

\def\Snospace~{\S{}}

\definecolor{Gray}{gray}{0.9}

\definecolor{csblue}{RGB}{68,119,170}
\definecolor{cscyan}{RGB}{102,204,238}
\definecolor{csgreen}{RGB}{34,136,51}
\definecolor{csyellow}{RGB}{204,187,68}
\definecolor{csred}{RGB}{238,102,119}
\definecolor{cspurple}{RGB}{170,51,119}
\definecolor{csgrey}{RGB}{187,187,187}

\newcommand{\ndcgat}[1]{\ensuremath{\mathrm{nDCG}{\scriptstyle @#1}}}
\newcommand{\dcgat}[1]{\ensuremath{\mathrm{DCG}{\scriptstyle @#1}}}
\newcommand{\cgat}[1]{\ensuremath{\mathrm{CG}{\scriptstyle @#1}}}
\newcommand{\psdcgat}[1]{\ensuremath{\mathrm{PSDCG}{\scriptstyle @#1}}}

\newtheorem{definition}{Definition}

\begin{document}
	\normalem %
\title{XFL: Naming Functions in Binaries with\\ Extreme Multi-label Learning}

\ifanonymous
\author{Anonymous Submission}
\else
\author{
  \IEEEauthorblockN{%
	James Patrick-Evans\IEEEauthorrefmark{1}\IEEEauthorrefmark{2},
	Moritz Dannehl\IEEEauthorrefmark{1}
	and Johannes Kinder\IEEEauthorrefmark{1}
	}
  \IEEEauthorblockA{%
    \IEEEauthorrefmark{1}%
	Research Institute CODE, Bundeswehr University Munich, Germany
  }
  \IEEEauthorblockA{%
    \IEEEauthorrefmark{2}%
    Information Security Group, Royal Holloway, University of London, United Kingdom\\
    james.patrick-evans.2015@rhul.ac.uk, \{moritz.dannehl,johannes.kinder\}@unibw.de
  }
}
\fi

\maketitle

\begin{abstract}
Reverse engineers benefit from the presence of identifiers such as function names in a binary, but usually these are removed for release. Training a machine learning model to predict function names automatically is promising but fundamentally hard: unlike words in natural language, most function names occur only once.
In this paper, we address this problem by introducing eXtreme Function Labeling (XFL), an extreme multi-label learning approach to selecting appropriate labels for binary functions. XFL splits function names into tokens, treating each as an informative label akin to the problem of tagging texts in natural language. 
We relate the semantics of binary code to labels through \Embeddings{}, a novel function embedding that combines static analysis-based features with local context from the call graph and global context from the entire binary. 
We demonstrate that XFL/\Embeddings{} outperforms the state of the art in function labeling on a dataset of 10,047 binaries from the Debian project, achieving a precision of 83.5\%.
We also study combinations of XFL with alternative binary embeddings from the literature and show that \Embeddings{} consistently performs best for this task. 
As a result, we demonstrate that binary function labeling can be effectively phrased in terms of multi-label learning, and that binary function embeddings benefit from including explicit semantic features.
\end{abstract}

\section{Introduction}

Software reverse engineering is the process of understanding the inner workings of a software system~\cite{ChikofskyC90}. 
In a computer security context, reverse engineering is typically performed on a binary without access to source code.
The goals of binary reverse engineering include security audits or vulnerability discovery~\cite{bughunting02}, interoperability with systems where no source code is available~\cite{Cifuentes99}, or forensic analysis of malicious code~\cite{sikorski-practicalmalware}.

Binary reverse engineering is a labor-intensive task, despite existing tooling to automate disassembly and function discovery. A particular challenge is the lack of readily available information about code. When binaries are released, helpful debugging information such as function and variable names are typically removed, or stripped, from the binary. This is done not only to reduce the size of the binary, but also to actively discourage reverse engineering of closed-source code and to protect intellectual property.

In an observational study, \citet{VotipkaRMFM20} identify three phases in reverse engineering: overview, sub-component scanning, and focused experimentation. The first two of these make heavy use of the available textual information in a binary, such as the names of called API functions or the contents of string constants. This suggests that additional textual information hinting at functionality in the binary would assist the reverse engineer at least during these initial phases that focus on mapping out the binary. Indeed, the study reports that most reverse engineers focus on improving readability of the code by adding their own annotations that essentially reconstruct debugging information like variable and function names or data structure types. Another recent study by Montavani et al.~\cite{Mantovani22} empirically confirms that especially experts frequently assign their own names to functions during reverse engineering.

Disassemblers such as IDA Pro, Ghidra, Radare2, and Binary Ninja have long identified this need and provide some automation to recognize and name common functions in frequently used software components. In addition to naming arguments to common API functions, they also perform binary pattern matching to name local functions from statically linked libraries and compiler runtimes. While this helpful functionality is widely-used to make reverse engineering more efficient, it is inherently limited. Signatures are designed to exactly match known functions, with only some flexibility to account for minor changes in compilation settings. 

Machine learning promises a new generation of more powerful tools for function identification, and initial academic work appears to confirm that it is possible to classify binary code into function names~\cite{debin,acsac20-punstrip}. These systems learn models of the contents and structure of functions and their most likely name. However, much of the success of these systems can be attributed to the identification of highly similar,
repeated functions across multiple binaries~(e.g., static library functions).

This type of approach faces two fundamental problems that limit its applicability: it can only generate function names that have been seen in the training set; and each such function name represents a separate output class, with the number of possible function names being essentially unbounded. Even worse, the classes are heavily imbalanced, with the majority of classes having a single sample and a minority of classes being over-represented (e.g., \verb|main|). Normalizing function names to remove differences in coding style such as \verb+CamelCase+ vs. \verb+snake_case+ can alleviate the problem to some extent~\cite{acsac20-punstrip}, but still no such approaches can accurately predict function names that remain unseen after normalization. 

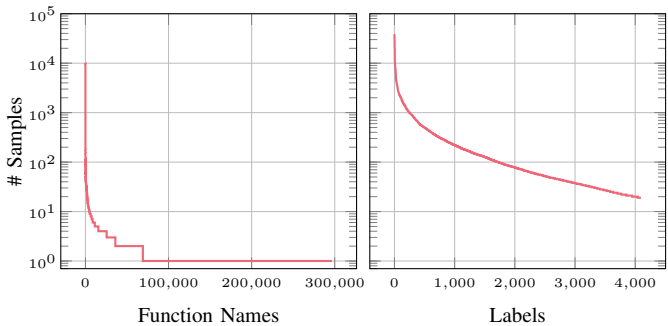
\begin{figure}[t]
\hspace*{-6pt}
\begin{tikzpicture}
	\tikzstyle{every node}=[font=\footnotesize]
	\pgfkeys{/pgf/number format/.cd,fixed}
	\begin{groupplot}[
		group style={
			group size=2 by 1,
			yticklabels at=edge left,
			ylabels at=edge left,
			horizontal sep=0.5em,
		},
		width=.444\columnwidth,
		scale only axis,
		scaled x ticks=false,
		ymode=log, 
		ylabel=\# Samples,
		ylabel shift=-6pt,
		yticklabel shift=-2pt,
		yticklabel style={font=\tiny},
		ytick distance=10,
		xticklabel style={font=\tiny},
		xmajorgrids,
		ymajorgrids,
		ymax=100000,
		ymin=0.7,
		]
		\nextgroupplot[
			xlabel=Function Names,
			xtick distance=100000,
			]
		\addplot [color=csred, thick, no markers] gnuplot [raw gnuplot] { plot "res/rnamefreq.txt" };
		
		\nextgroupplot[
			xlabel=Labels,
			xtick distance=1000,
			]
		\addplot [color=csred, thick, no markers] gnuplot [raw gnuplot] { plot "res/labelfreq.txt" };
	\end{groupplot}
\end{tikzpicture}
\caption{Semi-log plots showing the number of samples in each class when learning function names and labels, respectively. With whole function names, the vast majority of classes have only a single sample.}
\label{fig:dataset-sample-analysis}
\end{figure}

The drastic class imbalance is visible in \autoref{fig:dataset-sample-analysis}, which plots the frequencies of function names observed in a dataset of functions derived from 10,047 binaries in Debian packages. 
Six function names occur in at least 95\% of binaries (standardized names like \verb|main|, \verb|libc_csu_init|, etc.). Over 98\% of function names occur fewer than 10 times, and 73\% of function names occur only once.
This long tail of single-sample classes indicates that any model for whole function names is doomed to mispredict the vast majority of functions.

Our solution to this problem is to split function names into meaningful tokens. For instance, the \verb+xscreensaver+ function \verb+make_smooth_colormap+ would correspond to the set of labels $\{ \texttt{make}, \texttt{smooth}, \texttt{color}, \texttt{map}\}$. \autoref{fig:dataset-sample-analysis} shows that the imbalance in the label distribution is less pronounced. The total number of labels can be controlled such that each function has at least one descriptive label but there are also sufficiently many samples available per label. We therefore arrive at the problem of assigning a set of labels to each function.

A similar problem is that of tagging text with a set of relevant labels, which motivates multi-label learning~\cite{Joachims98,SchapireS00,NigamMTM00} and extreme multi-label learning (XML)~\cite{PrabhuV14,BhatiaJKVJ15}, where the number of labels is very large. Based on this insight, we show how to leverage state-of-the-art algorithms from XML for labeling functions in stripped binaries with \Tool{} (\emph{eXtreme Function Labeling}). \Tool{} scales to millions of data points and labels. 
\Tool{} is parameterized by a given \emph{function embedding}, which maps each binary function to a vector representation.
\Tool{} is compatible with state-of-the-art general-purpose binary code embeddings such as PalmTree~\cite{palmtree}, SAFE~\cite{safe} and Asm2Vec~\cite{asm2vec}. In addition, we designed and implemented the novel function embedding \Embeddings{}, which particularly emphasizes semantic properties of the binary code. To this end, \Embeddings{} is trained from a vector of per-function features combined with vectors capturing the context of the local call graph and of the whole binary. While this partly manual feature engineering runs counter to current trends in machine learning, we demonstrate that it is highly effective for the \Tool{} task, providing further evidence that semantic preprocessing of code can improve over syntactic language models~\cite{MukherjeeWCRCJ21}.

In summary, we make the following main contributions:

\begin{itemize}
    \item With \Tool{}, we introduce extreme multi-label learning as a solution to the problem of labeling binary functions~(\autoref{sec:xfl}). \Tool{} solves the problems of sparsity and class imbalance in binary function labeling and provides information-theoretic metrics for meaningful evaluation. In an extensive evaluation on a dataset with 741,724 functions from 10,047 binaries, we demonstrate that our implementation significantly outperforms the state of the art.

    \item We introduce \Embeddings{}~(\autoref{sec:embedding}), a new vector representation of binary functions using static analysis-based feature engineering and deep learning. We demonstrate that \Embeddings{} outperforms state-of-the-art function embeddings on the task of function labeling. This suggests that increasing the level of abstraction from assembly tokens to static analysis results improves the embedding quality.
    \item We present an end-to-end function name generation pipeline for stripped binaries based on \Embeddings{}, \Tool{}, and a language model for synthesizing plausible function names from assigned label sets~(\autoref{sec:ngrams-model}).

\end{itemize}

\section{Extreme Multi-label Learning}
\label{sec:background}

We now introduce multi-label classification~(\autoref{sec:background-xml}), review classic~(\ref{sec:multilabel-metrics}) and information gain-based~(\autoref{sec:measure-of-rank}) metrics, and introduce PfastreXML~(\autoref{sec:background-pfastrexml}), the state-of-the-art approach to extreme multi-label learning used in \Tool{}.

\subsection{Multi-Label Classification}
\label{sec:background-xml}

Multi-label classification is the problem of predicting a variably-sized set of labels per data point~\cite{Joachims98,SchapireS00,NigamMTM00}. This is different from multi-class classification, where each data point belongs to exactly one class. Because some labels can be more relevant to a data point than others, one also usually wishes to rank the labels by relevance.
In a small label space one can get away with a 1-vs-all approach and train an independent classifier for each label~\cite{AllweinSS00}. 

With larger label spaces, both training and prediction become too computationally expensive, however.
\textit{Extreme multi-label learning (XML)} deals with very large label spaces: the canonical XML problem  is that of predicting a set of suitable categories for a new Wikipedia page from the millions of categories of Wikipedia~\cite{PrabhuV14}. 
To make the problem tractable, both \textit{embedding-based} and \textit{tree-based} methods have been proposed in the literature. 

Embedding-based approaches aim to compress the label space by exploiting the fact that most data points will have only few labels, and that labels are highly correlated~\cite{TaiL12,CisseUAG13,YuJKD14,BhatiaJKVJ15}. While recent deep learning-based approaches appear promising~\cite{Dahiya21,slice,Mittal21}, embedding-based methods have traditionally been less precise than tree-based ones, on which we focus in this work.

Tree-based approaches combine classifiers in a hierarchy to reduce the number of classes in each individual problem. This can be done by partitioning either the label space~\cite{BengioWG10,DengSBL11} or the feature space. 
When partitioning the feature space~\cite{AgrawalGPV13,WestonMY13,PrabhuV14}, the feature space is split into regions such that in each such region only a small set of labels is active, i.e., has at least one training sample. Then a precise multi-label classifier can be trained for this reduced problem. 

Finding the partitioning is part of the training process; and as usual, the training process requires a loss function that should be minimized over the training samples. For the training process to be practical, the loss function should also be efficiently computable. 
In most XML settings, a given sample will have vastly more negative (irrelevant) labels than positive (relevant) ones. Therefore, the loss function should give more weight to correct positive labels than to correct negative labels. This rules out loss functions for smaller multi-label problems like Hamming distance~\cite{BalasubramanianL12}, which gives equal weight to positive and negative labels.
The key idea behind FastXML~\cite{PrabhuV14}, a precursor of the algorithm we use in this paper, is to directly optimize a loss function based on the \textit{rank} of labels, as in a \textit{learning to rank} problem, where labels are ranked by their relevance to the datapoint. 
FastXML uses the normalized discounted cumulative gain (nDCG), a metric introduced in information retrieval for calculating the usefulness of search results~\cite{JarvelinK02} (see \autoref{sec:measure-of-rank}).

\subsection{Multi-label Metrics}
\label{sec:multilabel-metrics}

Traditional metrics used in multi-label classification problems are precision, recall, and F-measure, in either their macro- or micro-average forms. \emph{Macro}-averaging computes a metric independently for each label and then averages across labels. This treats all labels equally, leading to skewed results when classes are highly imbalanced. In that situation, micro-averaging is preferred: \emph{micro}-averaging adds up true and false positives and negatives of all labels to compute an aggregate metric.
Because of the class-imbalance in function labels, we use micro-average scores throughout, in particular micro-average precision $P_{\mu}$, micro-average recall $R_{\mu}$, and micro-average $F_{1\mu}$ as defined below: 
\begin{gather*}
  P_{\mu}    = \frac{\sum_{\ell \in L}{{\mathit{TP}}_\ell}}{\sum_{\ell \in L}{{\mathit{TP}}_\ell + {\mathit{FP}}_\ell}} \quad
  R_{\mu}    = \frac{\sum_{\ell \in L}{{\mathit{TP}}_\ell}}{\sum_{\ell \in L}{{\mathit{TP}}_\ell + {\mathit{FN}}_\ell}}  \\[8pt]
  F_{1\mu}   = \frac{2 \cdot P_\mu \cdot R_\mu}{P_\mu + R_\mu}
\end{gather*}
Here, $L$ is the set of labels in the corresponding label space. $\mathit{TP_\ell}$, $\mathit{FP_\ell}$, $\mathit{FN_\ell}$, correspond to the number of true positives, false positives, and false negatives for each label defined in the standard way. %

Because they are widely-used and easy to interpret, we use micro-average precision, recall, and $F_1$ to evaluate \Tool{} on the task of predicting the relevant subset of correct labels assigned to each function name (\autoref{sec:evaluation-labeling}).
However, these metrics do not take the ranking information into account, which is why they are less suited for training an XML classifier. Therefore, FastXML and PfastreXML use the cumulative gain-based metrics below; we also use these metrics to give more insight into the effectiveness of \Embeddings{} and \Tool{} in our evaluation.

\subsection{Cumulative Gain-based Metrics}
\label{sec:measure-of-rank}

Cumulative gain-based metrics are widely used in information retrieval~\cite{BhatiaJKVJ15, AgrawalGPV13, HsuKLZ09, PrabhuV14, WestonBU11, WestonMY13, YuJKD14}. Intuitively, they sum up the gain in useful information from looking at the first $k$ results of a search query~\cite{JarvelinK02}. This makes them ideal for evaluating the quality of a ranked list of results, or equivalently, the relevance of the ranked labels for a datapoint.
\begin{definition}[Cumulative Gain]
  The cumulative gain $\cgat{k}$ measures the information gain in the top $k$ labels of a ranked list of labels and is defined as
  \[ \cgat{k} = \sum^{k}_{i=1} \mathit{rel}_i \]
  where $\mathit{rel}_i$ is the relevance of the label at position $i$. 
\end{definition}

In this work, we use a binary relevance of either 1 or 0, depending on whether the label is contained in the set of true labels for each data point or not. 
Note that the cumulative gain ignores the ordering of labels within the top $k$ elements, so for our purposes it simply counts the number of correct labels.

The \emph{discounted cumulative gain} introduces a logarithmic discounting factor to the relevance of each label such that labels appearing earlier are given more weight:
\begin{definition}[Discounted Cumulative Gain]
  \[ \dcgat{k} = \sum^{k}_{i=1} \frac{ {rel}_i }{ \log_2 (i+1) } \]
\end{definition}

Neither cumulative gain nor discounted cumulative gain account for the difference in the true relevance of labels for each data point or query: some may have many relevant labels, others only few. To be comparable across data points, the \emph{normalized discounted cumulative gain} (nDCG) multiplies the DCG with a normalization factor $Z_k$ that is the inverse of the maximum DCG achievable by returning the $k$ truly most relevant labels. 
In our case of binary relevance, we thus have 
\[ Z_k = \left( \sum_{i=1}^{\min(k,n)} \frac{1}{\log_2 (i+1)} \right)^{-1} \] 
for a data point with $n$ true labels, as all other labels will have a relevance of 0. 
We can now define:
\begin{definition}[Normalized Discounted Cumulative Gain]
  \[ \ndcgat{k} = Z_k \cdot \sum^{k}_{i=1} \frac{ {rel}_i }{ \log_2 (i+1) } \]
\end{definition}
The nDCG produces a metric for evaluating multi-label classification in the range $[0,1]$ whereby a perfect ranking would achieve a score of $1.0$.
To evaluate over multiple points in a dataset, we compute the mean of all nDCG scores.

In our evaluation, we consistently use $k=5$ as most functions will tokenize to five or fewer labels in the label-space sizes we considered (see also the dataset analysis in~\autoref{tab:dataset-analysis}).

\subsection{Propensity-based Scoring in PfastreXML}
\label{sec:background-pfastrexml}

A common issue in the very large datasets used for training XML models is that data points in the observed ground truth often do not carry all truly relevant labels (the complete ground truth). In datasets like Wikipedia, authors and editors may not simply not be aware of all the categories that would in principle apply to their article. 
It is much less common to have labels assigned to data points that are plain wrong, however. The noise in the ground truth therefore mostly goes into the direction of data points missing labels.

A key insight behind \textit{PfastreXML}~\cite{JainPV16} is that successful predictions of labels which are rarely assigned although they are relevant, should be specifically rewarded during training.
To that end, PfastreXML improves FastXML by using label \textit{propensities}~\cite{RosenbaumR83} as part of the loss function. 
Intuitively, for a specific data point the propensity of a label is the probability that it has actually been assigned in the ground truth if it is truly relevant to the data point.
The propensity-scored loss function then estimates the true loss that would be observed on the complete ground truth without any missing labels. 

Based on empirical evidence on large datasets that permit reliable estimates of the true ground truth, Jain et al.~\cite{JainPV16} propose that the propensity $p_\ell$ of label $\ell$ can be modeled by a sigmoidal function of $\log N_\ell$, where $N_\ell$ is the number of data points annotated with label $\ell$ in the observed ground truth dataset of size $N$:
\begin{equation*}
    \label{eq:pfastrexml-propensities}
    p_\ell = \frac{1}{1 + C e^{-A \log ( N_\ell + B )}}
\end{equation*}
Here, $A$ and $B$ are hyper-parameters, and $C$ is derived from them as $C \!=\! (\log N - 1 ){(B + 1)}^A$. The hyper-parameters should be tuned for each dataset such that plotting the propensities of all labels against their respective $\log N_\ell$ fits a sigmoid.

PfastreXML uses propensities by defining a propensity-scored variant PSDCG of the DCG for computing its loss function, in particular it multiplies the discount with the corresponding label propensity such that 
	\[ \psdcgat{k} = \sum^{k}_{i=1} \frac{ {rel}_i }{ p_{\ell_i} \log_2 (i+1) } \]
where $\ell_i$ is the label predicted at rank $i$.

Finally, PfastreXML also adjusts the final classification step in the leaf nodes of the tree learned by FastXML by reranking results using classifiers for rare labels, which introduce additional hyperparameters $\alpha$ (for re-ranking) and $\gamma$ (for predicting rare labels).
Through propensity-scoring and reranking, PfastreXML improves the multi-label classification accuracy of FastXML especially when dealing with infrequent labels in large datasets.

\tikzexternaldisable
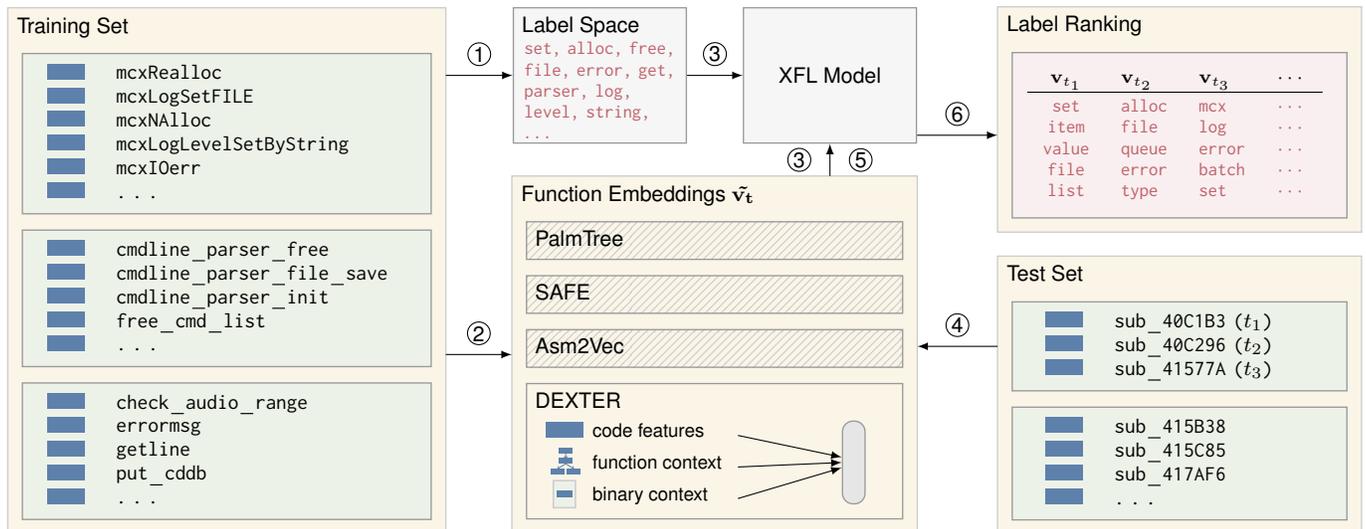
\begin{figure*}[t]
	\tikzset{external/optimize=false}

\definecolor{blue}{HTML}{5e81ac}
\definecolor{green}{HTML}{a3be8c}
\definecolor{red}{HTML}{bf616a}
\definecolor{darkblue}{HTML}{000000}
\definecolor{yellow}{HTML}{ebcb8b}

\newcommand{\GENERICCFG}{
	\scalebox{0.35}{
		\tikzset{external/optimize=false}
		\begin{tikzpicture}[inner sep=0mm,
			level 1/.style={sibling distance=6mm,level distance=4mm},
			level 2/.style={sibling distance=6mm,level distance=4mm},
			level 3/.style={sibling distance=6mm,level distance=4mm},
			cfgnode/.style={fill=blue, minimum width=5mm, minimum height=2mm},
			edge from parent path={(\tikzparentnode.south) -- (\tikzchildnode.north)},
			thin, ->, blue
			]
			
			\node[cfgnode] (A) {}
			child {
				node[cfgnode] (B) {}
				child {
					node[cfgnode] (C) {}
				}
				child {
					node[cfgnode] (D) {}
				}
			};
			\path let \p1 = (C.west) in node[] (IGW) at (\x1-1mm,\y1) {};
		\end{tikzpicture}
	}
}

\newsavebox{\INITIALCFG}
\begin{lrbox}{\INITIALCFG}
	\tikzset{external/optimize=false}
	\hspace*{-1pt}\GENERICCFG{}
\end{lrbox}

\newcommand{\tripleblue}{
\raisebox{-3pt}{\scalebox{.4}{%
	\begin{tikzpicture}
		\node[draw=black!20, fill=green!20, line width=.5mm, minimum width=.3cm, centered, minimum height=.9cm] (x) {\bluebox};
	\end{tikzpicture}}}}

\tikzset{>=latex}
\def\blockdist{3.5}
\def\edgedist{2.5}
\pgfdeclarelayer{background}
\pgfdeclarelayer{foreground}
\pgfsetlayers{background,main,foreground}
\tikzset{external/optimize=false}
\begin{tikzpicture}
	[
	x=6.2cm,
	y=3cm,
	node distance=2mm and 2cm,
	box/.style={draw=black!40, line width=.5pt, align=left, rectangle, minimum width=5cm},
	sbox/.style={box, minimum height=5mm, pattern=north east lines, pattern color=black!20},
	func/.style={box, fill=green!20, line width=.5pt, text width=5.2cm, align=left, font=\footnotesize\ttfamily},
	lab/.style={line width=.5pt, font=\scriptsize\ttfamily,text=red},
	grouppanel/.style={box,draw=black!20,fill=yellow!20},
	]
	
\node[func] at (0,-0.2) (bot) {
\begin{tabular}{c l}
	\bluebox & check\_audio\_range\\
	\bluebox & errormsg\\
	\bluebox & getline\\
	\bluebox & put\_cddb\\
	\bluebox & \ldots\\
\end{tabular}
	};
\node[func] (mid) [above=of bot] {
\begin{tabular}{c l}
	\bluebox & cmdline\_parser\_free \\
	\bluebox & cmdline\_parser\_file\_save \\
	\bluebox & cmdline\_parser\_init\\
	\bluebox & free\_cmd\_list\\
	\bluebox & \ldots\\
\end{tabular}
};
\node[func] (top) [above=of mid] {
\begin{tabular}{c l}
	\bluebox & mcxRealloc \\
	\bluebox & mcxLogSetFILE \\
	\bluebox & mcxNAlloc  \\
	\bluebox & mcxLogLevelSetByString \\
	\bluebox & mcxIOerr \\
	\bluebox & \ldots\\
\end{tabular}
};

\node[box, align=left] at (1.046,-6mm) (dexter) {\\[3mm]
\hspace*{-24mm}
\begin{tabular}{c@{}l}
	\bluebox & \sffamily\scriptsize code features\\
	\hspace*{-4pt}\raisebox{-2pt}{\usebox{\INITIALCFG}} & \sffamily\scriptsize function context\\
	\tripleblue & \sffamily\scriptsize binary context
\end{tabular}};
\node[below right,font=\footnotesize\sffamily] at (dexter.north west) {DEXTER};
\node[left, draw=black!40, line width=.5pt, align=left, rounded corners=4pt, xshift=-5mm, yshift=-1.5mm, minimum height=11mm, minimum width=3mm, fill=gray!20] at (dexter.east) (dexvec) {};

\path[draw, ->] (dexter.center)+(0.05,0.07) -- ([yshift=2pt]dexvec.west);
\path[draw, ->] (dexter.center)+(0.05,-0.07) -- (dexvec.west);
\path[draw, ->] (dexter.center)+(0.05,-0.21) -- ([yshift=-2pt]dexvec.west);

\node[sbox] (asm2vec) [above=of dexter] {};
\node[below right,font=\footnotesize\sffamily] at (asm2vec.north west) {Asm2Vec};

\node[sbox] (safe) [above=of asm2vec]{};
\node[below right,font=\footnotesize\sffamily] at (safe.north west) {SAFE};

\node[sbox] (palmtree) [above=of safe]{};
\node[below right,font=\footnotesize\sffamily] at (palmtree.north west) {PalmTree};

\node[box, fill=gray!7, minimum height=18mm, minimum width=2.3cm] at (.8, 44mm) (labelspace) {};
\node[below right,font=\footnotesize\sffamily] at (labelspace.north west) {Label Space};
	
\node[below=2mm, lab, text width=20.1mm, minimum height=18mm] at (labelspace.north) (l1) {set, alloc, free, file, error, get, parser, log, level, string, $\ldots$};

\node[box, fill=gray!7, minimum height=18mm, minimum width=2.3cm, font=\footnotesize\sffamily, right=19mm] at (labelspace) (xfl) {XFL Model};

\node[func, text width=4cm, minimum width=4.45cm] at (2.043,0.265) (target) {
\begin{tabular}{c l}
	\bluebox & \textcolor{black}{sub\_40C1B3} ($t_1$) \\
	\bluebox & \textcolor{black}{sub\_40C296} ($t_2$)\\
	\bluebox & \textcolor{black}{sub\_41577A} ($t_3$)\\
\end{tabular}
};
\node[func, text width=4cm, minimum width=4.45cm] (target2) [below=of target] {
\begin{tabular}{c l}
	\bluebox & sub\_415B38 \\
	\bluebox & sub\_415C85 \\
	\bluebox & sub\_417AF6 \\
	\bluebox & \ldots\\
\end{tabular}
};
	
\node[box, fill=red!10, font=\scriptsize\ttfamily, minimum height=2.2cm, minimum width=4.45cm, above=17mm]  at (target) (lrank) {
	\begin{tabular}{c  l l l }
		$\color{darkblue}\mathbf{v}_{t_1}$ & $\color{darkblue} \mathbf{v}_{t_2}$ & $\color{darkblue}\mathbf{v}_{t_3}$ & $\color{darkblue}\cdots$\\[1pt]
		\hline & & & \\[-2ex]
		\color{red} set & \color{red} alloc & \color{red} mcx & \color{red}$\cdots$\\
		\color{red} item & \color{red} file  & \color{red} log& $\color{red} \cdots$\\
		\color{red} value & \color{red} queue  & \color{red} error& $\color{red} \cdots$\\
		\color{red} file & \color{red} error  & \color{red} batch& $\color{red} \cdots$\\%
		\color{red} list & \color{red} type & \color{red} set & $\color{red} \cdots$
	\end{tabular}
};

\begin{pgfonlayer}{background}
	\path (top.west |- top.north)+(-0.03,0.2) node (a) {};
	\path (bot.east |- bot.south)+(0.03,-0.06) node (b) {};
	\path[grouppanel]
	(a) rectangle (b);
	
	\path (target.north west)+(-0.03,0.2) node (c) {};
	\path (target2.south east)+(0.03,-0.06) node (d) {};
	\path[grouppanel]
	(c) rectangle (d);

	\path (palmtree.north west)+(-0.03,0.2) node (e) {};
	\path (dexter.south east)+(0.03,-0.06) node (f) {};
	\path[grouppanel]
	(e) rectangle (f);

	\path (lrank.north west)+(-0.03,0.2) node (g) {};
	\path (lrank.south east)+(0.03,-0.06) node (h) {};
	\path[grouppanel]
	(g) rectangle (h);

\end{pgfonlayer}

\node[below right,font=\footnotesize\sffamily] at (a) {Training Set};
\node[below right,font=\footnotesize\sffamily] at (c) {Test Set};
\node[below right,font=\footnotesize\sffamily] at (e) {Function Embeddings \color{darkblue}$\mathbf{\vec{v_t}}$};
\node[below right,font=\footnotesize\sffamily] at (g) {Label Ranking};

\node[] at ($ (e)!0.5!(f) $) (z) {};

\path[draw, ->] (labelspace.east) -- node [above] {\circled{3}} (xfl.west);
\path[draw, ->] (xfl.east |- lrank.west) -- node [above] {\circled{6}} (g |- lrank.west);
\path[draw, ->] (b |- z) -- node [above] {\circled{2}} (e |- z);
\path[draw, ->] (b |- labelspace.west) -- node [above] {\circled{1}} (labelspace.west);
\path[draw, ->] (xfl.south |- e) -- node {\circled{3}\hspace*{5mm}\circled{5}} (xfl.south);
\path[draw, ->] (c |- target.west) -- node [above] {\circled{4}} (f |- target.west);

\end{tikzpicture} 	%
	\caption{Overview of the \Tool{} training and inference process. Function names in the training dataset are preprocessed to create the label space \protect\circled{1}. The function bodies are used to train the embeddings \protect\circled{2}. Function embeddings and the label space serve as input for training an \Tool{} model \protect\circled{3}. To infer labels for functions in a binary, an embedding $\mathbf{v}_{t_i}$ is calculated for each unknown function $t_i$ \protect\circled{4} and fed into the \Tool{} model \protect\circled{5}, which then produces a ranked list of labels per function \protect\circled{6}.}
	\label{fig:overview}
\end{figure*}
\tikzexternalenable

\section{Overview}
\label{sec:overview}

We now give a brief overview of our proposed end-to-end architecture for function labeling, including the processes for training~(\autoref{ref:overview-training}) and prediction (\autoref{sec:overview-prediction}). The architecture is shown in \autoref{fig:overview}.
When given a stripped binary, our system is trained to predict a ranked set of labels for each function, corresponding to tokens found in the names of similar functions. These labels can inform a reverse engineer directly in their ranked form, or they can be used to automatically synthesize a likely function name containing them. 

\subsection{Training}
\label{ref:overview-training}

There are several components that require training. Training data consists of unstripped binary executables, such that we know the developer-assigned names of functions. In a practical deployment, one would collect as many binaries as possible; for evaluating performance in our paper, we split the available data into training, validation, and test sets.

\subsubsection{Generating a Label Space}
\label{ref:overview-labelspace}
We need to define a label space from which we will draw the labels to predict. To that end, we tokenize all function names in the training set according to a set of syntactic rules. The rules take into account (combinations of) multiple naming conventions and substitute common abbreviations. The union of all tokens becomes the label space. We bound the size of the label space to between 512 and 4096 labels in our experiments, which excludes extremely rare and unique tokens such as typos, highly program-specific or non-English words.

\subsubsection{Training the Function Embeddings}
\label{ref:overview-embeddings}
Binary function embeddings are a fundamental building block for our approach. They map the code of a function within a binary to a vector representation. The training process for function embeddings attempts to optimize the representation such that similar functions have low distance in the vector space, whereas dissimilar functions are further apart, for some notion of similarity. We can either use (partially) pre-trained embeddings from the literature, or our own embedding, \Embeddings{}. 

\subsubsection{Training \Tool{}}
\label{ref:overview-xfl}
We train the extreme multi-label classifier to predict sets of labels for each embedding vector, based on the tokenized name of its corresponding binary function. PfastreXML maximizes the nDCG for the training data, i.e., it aims to ensure that the labels in a function's name will be ranked highest for that data point among all the labels in the label space. The observed probabilities of labels for training samples also imply a threshold value for the probabilities of true labels in a ranking of all labels.
To improve performance, one can perform hyper-parameter tuning as part of this step, in particular for parameters $A$, $B$, $\alpha$, and $\gamma$.

\subsubsection{Training the Language Model}
\label{ref:overview-ngrams}
To generate actual function names, we train a classical trigram language model from the tokenization of function names in the training set. From a set of labels, the language model is then able to predict their most likely order in a real function name.

\subsection{Prediction}
\label{sec:overview-prediction}
Our resulting system will predict labels for stripped binaries, as a reverse engineer would encounter them. 
We convert each target binary function into a vector using the trained embeddings model. We feed that vector into the \Tool{} model, which will produce a ranking for all labels in the label space. All labels with probabilities above the threshold value observed during training are returned as the predicted set.
From the predicted set of labels, the language model can then synthesize the most likely function name containing all these labels according to a coding convention like \verb+snake_case+.

\section{Function Embedding}
\label{sec:embedding}

We now present our approach to function embeddings in detail. First, we briefly describe existing embeddings for binary functions~(\autoref{sec:embeddings-soa}), before introducing the features of our new \Embeddings{} embedding~(\autoref{sec:feature-engineering}), its representation of context~(\autoref{sec:function-context}), and the training process~(\autoref{sec:autoencoder}).

\subsection{Existing Binary Function Embeddings}
\label{sec:embeddings-soa}

Function embeddings represent binary code with a real-valued vector, capturing similarities and arranging the vector space such that (syntactically or semantically) similar functions have a smaller distance between them relative to other dissimilar functions.  We discuss three existing embeddings for binary code, Asm2Vec~\cite{asm2vec}, SAFE~\cite{safe}, and PalmTree~\cite{palmtree}, all of which can be used with \Tool{}. All three are inspired by natural language processing, treating instructions sequences and possibly additional information as sentences.  %

Asm2Vec~\cite{asm2vec} adopts a Word2Vec-like approach~\cite{word2vec} for assembly language. For every assembly function, the model generates execution traces and applies a Paragraph Vector Distributed Memory~(PV-DM) model on it, generating a distributed representation for  opcodes and operands of assembly instructions. Along the way, a vector representation for assembly functions is learned, similar to paragraph vectors~\cite{paragraph2vec}.

SAFE~\cite{safe} uses a self-attentive neural network architecture and models sequences of assembly code. SAFE first models each instruction using an adapted skip-gram method. The sequence of instruction embeddings is then used to compute local summaries using a self-attentive neural network, and the summaries are combined in a weighted sum. 

PalmTree~\cite{palmtree} is a BERT-based~\cite{devlin-bert} model generating instruction embeddings from assembly code. It is trained via self-supervised learning using multiple training objectives on assembly code, such as Masked Language Model, Next Sentence Prediction, and Def-Use Prediction.
To obtain function embeddings from PalmTree, the authors use the Gemini Siamese architecture~\cite{gemini}.

Both Asm2Vec and SAFE are syntactic in nature, although they use static analysis for constructing the CFG and retrieving feasible traces of assembly instructions. PalmTree adds def-use information, which makes results of static analysis available for training the embedding.

\subsection{\Embeddings{} Feature Engineering}
\label{sec:feature-engineering}

Our hypothesis behind the \Embeddings{} embeddings is that making code semantics explicit will help the training process derive meaningful embeddings with less training data. Where existing embeddings largely follow a modern natural language processing pipeline that does only minimal preprocessing~\cite{asm2vec,safe,palmtree}, we consciously go against current trends in deep learning and follow an approach that gives more weight to classical feature engineering.

\begin{table}[t]
    \caption{Binary code features used by \Embeddings{} embeddings.}
\label{tab:symbol-features}
\centering
\begin{tabularx}{\linewidth}{ X }
    \toprule
	\multicolumn{1}{c}{ \textit{Quantitative Features}: $\mathbf{q}$ } \\
	\midrule
    Size of the symbol in bytes \\ 
    Number of IR instructions \\ 
	Sum of one-hot-encoded vectors of branch types \\ 
    Number of temporary variables in the IR \\ 
	Sum of one-hot-encoded vectors of IR elements \\ 
    Number of callers \\
    Number of callees \\
	Number of transitively reachable functions \\ 
	Vector representation of the function CFG \\
	Vector representation of the function node in the binary callgraph \\
	Number of bytes referenced on the stack \\
	Number of bytes referenced on the heap \\
    Number of bytes referenced in Thread Local Storage \\
    Number of function arguments \\ 
	Number of bytes used for local variables on the stack \\
	One-hot encoded vector of tainted register types \\
	Number of tainted bytes of the heap  \\
	Number of tainted bytes of the stack  \\
	Number of tainted bytes in arguments to other functions \\
	Number of conditional jumps that depend on a tainted variable \\
    Number of tainted flows to other functions  \\
	\midrule
	\multicolumn{1}{c}{ \textit{Categorical Features}: $\mathbf{c}$ } \\
    \midrule
    Common SHA-256 hashes of assembly opcodes \\ 
    Common MinHash hashes of assembly opcodes \\
    Common constants referenced \\ 
    Names of dynamically linked callees   \\
    Known function names reachable from this function \\ 
    References to known data objects in dynamically linked libraries \\
    Names of dynamic functions and argument registers tainted \\
    \bottomrule
\end{tabularx}
\end{table}

Our features build upon those derived by the analysis framework of \citet{acsac20-punstrip}, which 
lifts each function to an intermediate representation (VEX) and performs a symbolic analysis of each function in isolation. 
We analyze live memory addresses and register values used in each function to determine the number of input arguments. Finally, we taint each input argument recovered and
calculate flows to the individual arguments of callee functions. The full list of features is shown in~\autoref{tab:symbol-features}.
We use high-level intra-procedural and inter-procedural features in order to mitigate large differences in machine code from different compilation environments. Our intuition for using these features comes from Kim et al.'s overview of binary similarity~\cite{bin-sim-lessons-learned} which analyzes the most prevalent features for matching similar binary functions.

We add two graph based vector representations of each function. The first uses the Local Degree Profile~(LDP)~\cite{LocalDegreeProfile-graph-embedding} graph embedding technique to map each function's intra-procedural CFG into a vector representation. The second uses uses the BoostNE~\cite{BoostNE} node embedding algorithm to embed the function's location in the callgraph.

Our features also include two hashes of assembly opcode sequences to aid the identification of common functions: a SHA-256 hash matches exact opcode sequences, and the locality-sensitive hashing algorithm MinHash~\cite{minhash} matches very similar opcode sequences.

The feature vector for a binary function $b$ is built from concatenating two vectors of quantitative and categorical data: 
\[
    \mathbf{f}_b = \left [ \mathbf{q}_b , \mathbf{c}_b \right ]
\]
Here, $\mathbf{q}$ represents dense, quantitative information describing features of functions in executables and $\mathbf{c}$ represents sparse, categorical data of features that have no numerical basis, in a one-hot-encoded form.
Each quantitative feature vector contains 512 elements, and each categorical feature vector contains 3~million elements. 

\subsection{Function and Binary Context}
\label{sec:function-context}

In addition to features of the function itself, we include information from its local and global context in \Embeddings{}. 
To represent the calling context, we build a function context vector $\mathbf{f}_C$ as the mean of the feature vectors of its callers and callees. This captures information from surrounding functions where the body of the target function does not contain enough unique information to correctly predict its name.
For example, overloaded functions, which can be called with a different number of parameters, will typically have small stub functions that initialize parameters and then jump to the larger function implementing the actual functionality.
We define the function context of a function $b$ as
\[
	\mathbf{g}_b =  \frac{1}{\left | \mathcal{C}_b \right | } \sum_{k \in \mathcal{C}_b} \mathbf{f}_{k} 
\]
where $\mathcal{C}_b$ is the set of callers and callees of function $b$.

We also wish to include context from the entire binary, as especially the naming of functions can be affected by the context in which it is found (e.g., some libraries use common prefixes to all their functions). To this end, we average the feature vector of every function in the whole binary and define the binary context as
\[
	\mathbf{h}_b = \frac{1}{\left | \mathcal{B}_b \right | } \sum_{k \in \mathcal{B}_b} \mathbf{f}_{k} 
\]
where $\mathcal{B}_b$ is the set of all functions in the binary that $b$ is in.

Finally, we build a modified feature vector $\hat{\mathbf{f}}_b$ for each function $b$ in our dataset by concatenating the function feature vector with its context:
\[
    \hat{\mathbf{f}}_b = \left [ \mathbf{f}_b , \mathbf{g}_b , \mathbf{h}_b \right ]
\]
This modified feature vector of each function is then used as input to train an autoencoder.

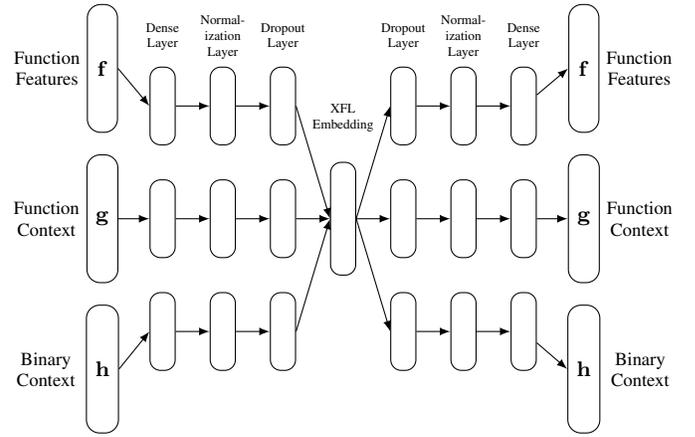
\begin{figure}[t]
\tikzset{>=latex}%
\hspace*{-1mm}
\begin{tikzpicture}
	[
	x=8mm,
	node distance=.4cm and 5.5cm,
	descr/.style={align=center, rectangle, 
		font=\tiny, 
		text width=1cm,
		inner sep=0pt,
		outer sep=0pt},
	odescr/.style={descr, font=\scriptsize},
	layer/.style={draw, minimum width=4mm, minimum height=17mm, rounded corners=5pt, font=\footnotesize, align=center},
	slayer/.style={draw, text width=.1cm, text height=.8cm, rounded corners},
	]
	\node[layer] at (0,5) (lay1) {$\mathbf{f}$};
	\node[layer] at (0,3) (lay2) {$\mathbf{g}$};
	\node[layer] at (0,1) (lay3) {$\mathbf{h}$};
	
	\node[odescr, left=.3mm of lay1] (f) {Function Features};
	\node[odescr, left=.3mm of lay2] (fc) {Function Context};
	\node[odescr, left=.3mm of lay3] (bc) {Binary Context};
	
	\node[slayer] at (1, 4.5) (dlay1) {};
	\node[slayer] at (1, 3) (dlay2) {};
	\node[slayer] at (1, 1.5) (dlay3) {};
	\node[descr, above=of dlay1, centered, align=center] {Dense\\Layer};
	
	\node[slayer] at (2, 4.5) (nlay1) {};
	\node[slayer] at (2, 3) (nlay2) {};
	\node[slayer] at (2, 1.5) (nlay3) {};
	\node[descr, above=of nlay1, centered, align=center, text width=1.0cm] {Normal\-ization\\Layer};
	
	\node[slayer] at (3, 4.5) (drlay1) {};
	\node[slayer] at (3, 3) (drlay2) {};
	\node[slayer] at (3, 1.5) (drlay3) {};
	\node[descr, above=of drlay1, centered, align=center] {Dropout\\Layer};
	
	\node[slayer, minimum height=15mm] at (4,3) (xfllayer) {};
	\node[descr, above=of xfllayer, text width=1.6cm, align=center] {XFL\\Embedding};

	\node[slayer] at (5, 4.5) (dlay11) {};
	\node[slayer] at (5, 3) (dlay12) {};
	\node[slayer] at (5, 1.5) (dlay13) {};
	\node[descr, above=of dlay11, centered, align=center] {Dropout Layer};
	
	\node[slayer] at (6, 4.5) (nlay11) {};
	\node[slayer] at (6, 3) (nlay12) {};
	\node[slayer] at (6, 1.5) (nlay13) {};
	\node[descr, above=of nlay11, centered, align=center, text width=1.0cm] {Normal\-ization\\Layer};
	
	\node[slayer] at (7, 4.5) (drlay11) {};
	\node[slayer] at (7, 3) (drlay12) {};
	\node[slayer] at (7, 1.5) (drlay13) {};
	\node[descr, above=of drlay11, centered, align=center] {Dense\\Layer};
	
	\node[layer] at (8,5) (lay11) {$\mathbf{f}$};
	\node[layer] at (8,3) (lay12) {$\mathbf{g}$};
	\node[layer] at (8,1) (lay13) {$\mathbf{h}$};
	
	\node[odescr, right=.3mm of lay11] (f1) {Function Features};
	\node[odescr, right=.3mm of lay12] (fc1) {Function Context};
	\node[odescr, right=.3mm of lay13] (bc1) {Binary Context};

	\draw[->] (lay1.east) -- (dlay1.west);
	\draw[->] (lay2.east) -- (dlay2.west);
	\draw[->] (lay3.east) -- (dlay3.west);
	
	\draw[->] (dlay1) -- (nlay1);
	\draw[->] (dlay2) -- (nlay2);
	\draw[->] (dlay3) -- (nlay3);
	
	\draw[->] (nlay1) -- (drlay1);
	\draw[->] (nlay2) -- (drlay2);
	\draw[->] (nlay3) -- (drlay3);
	
	\draw[->] (drlay1.east) -- (xfllayer.west);
	\draw[->] (drlay2.east) -- (xfllayer);
	\draw[->] (drlay3.east) -- (xfllayer.west);
	
	\draw[->] (xfllayer.east) -- (dlay11.west);
	\draw[->] (xfllayer) -- (dlay12.west);
	\draw[->] (xfllayer.east) -- (dlay13.west);
	
	\draw[->] (dlay11) -- (nlay11);
	\draw[->] (dlay12) -- (nlay12);
	\draw[->] (dlay13) -- (nlay13);
	
	\draw[->] (nlay11) -- (drlay11);
	\draw[->] (nlay12) -- (drlay12);
	\draw[->] (nlay13) -- (drlay13);
	
	\draw[->] (drlay11) -- (lay11.west);
	\draw[->] (drlay12) -- (lay12.west);
	\draw[->] (drlay13) -- (lay13.west);

\end{tikzpicture} 	\caption{Overview of the autoencoder design used to generate \Embeddings{} embeddings.}
	\label{fig:autoencoder-diagram}
\end{figure}

\subsection{Autoencoder Training}
\label{sec:autoencoder}

A deep autoencoder is then trained on our modified feature vectors for each function to create a dense, distributed embedding of functions.
The autoencoder architecture is depicted in~\autoref{fig:autoencoder-diagram}. The model first creates dense representations for the function and its contexts before combining them into a single embedding.
Our methodology captures structural information not present when using features from each function in isolation. 
To enforce generalization we first connect the input layer into three dense sub-layers, each with 768 nodes, before performing batch normalization and creating a dropout layer. 
These layers, along with $L_1$ and $L_2$ regularization on each dense sub-layer, aim to prevent our model from over-fitting and force our model to learn an embedding that generalizes well. 
Finally, we connect the output from all three dropout layers into a 512 node dense layer that we use as our embedded representation. 
The model architecture is reversed for predicting the output feature vector, with the exception that, 
during training we apply Gaussian noise immediately after the embedding layer using standard deviation of 0.1.

Our model is created using Tensorflow~\cite{tensorflow2015-whitepaper} and aims to minimize the loss between an input feature vector and a corresponding output feature vector. 
When feeding our model pairs of input and output vectors, the output feature vector is randomly sampled from the set of all functions with the same name as the corresponding input feature vector.
To effectively train useful embeddings, we optimize our model using max margin contrastive loss with Adagrad. 
Using standard loss metrics such as the Binary Cross Entropy or Hamming loss is ineffective at training models with very sparse data.
In using the contrastive loss, we aim to produce an embedding in which functions with similar features appear close together in the embedded space.

\section{XML for Function Binaries}
\label{sec:xfl}

\Tool{} uses PfastreXML and a binary function embedding to efficiently perform multi-label classification for tokenized labels of function names. 
We first detail how \Tool{} splits function names into tokens~(\autoref{sec:label-tokenizing}) and generates a label space~(\autoref{sec:label-space}), before describing how \Tool{} uses 
PfastreXML to rank associated labels~(\autoref{sec:label-prediction}) for functions.

\subsection{Tokenizing Function Names}
\label{sec:label-tokenizing}

For \Tool{} to predict labels in function names we first need a well defined label space consisting of string tokens found in the names of functions.
Labels should be informative, take into account programming styles and be mutually exclusive where possible; for example, the tokens \verb+str+, \verb+string+, \verb+String+, and \verb+__xStr__+ should all have the common denominator token \verb+string+.
\Tool{} generates a \emph{canonical token set} for each function name, a set of string tokens that canonically describe it. We generate a well-defined label space of a fixed size by analyzing the union of the resulting canonical token sets from the corpus of function names in the training set.

To generate the canonical token set $L_c$ from a function name, \Tool{} uses the following procedure:

\subsubsection{Strip Library Decorations} Regular expressions remove common symbol annotations added by compilers, e.g., \verb|'.*\.constp$'|, \verb|'.^\.avx\d+'|. Regular expressions for Radare2, IDA Pro, and Ghidra annotations are also applied depending on the analysis platform.
\subsubsection{Split Alphanumerical} The function name is split into character sequences along non-alphanumeric characters. Numeric and alpha characters are further split into separate groups.
    For example, $\texttt{\_\_libxyz\_init} \mapsto \left \{ \texttt{libxyz} , \texttt{init} \right \}$.
\subsubsection{Split Camel Case} We recognize common naming conventions in C binaries and split a  continuous character sequence into sets if we detect the use of camel case. For example, $\texttt{IsWindowOpen} \mapsto \left \{ \texttt{is} , \texttt{window} , \texttt{open}  \right \}$.
\subsubsection{Abbreviation Expansion} We expand a predefined list of common programming abbreviations such as \verb|fd| for \textit{file descriptor} and \verb|init| for \textit{initialization}. For example, $\texttt{mkdirs} \mapsto \left \{ \texttt{make}, \texttt{directories} \right \}$.
\subsubsection{Best Split of the Rod} We use a dynamic programming algorithm to split character sequences into the largest possible non-overlapping sequences. We check all permutations of sub-sequences to find the largest collection of English words. Our algorithm scores longer words higher over two or more same-length sub-sequences, e.g., $\left \{ \texttt{background} \right \} > \left \{ \texttt{back}, \texttt{ground} \right \}$. However, a longer length of total characters scores higher, e.g., $\texttt{foreach} \mapsto \left \{ \texttt{for}, \texttt{each} \right \}$ and not $\left \{ \texttt{reach} \right \}$.

\smallskip
To confirm that our method produces tokens representative of the developer intent, we conducted a manual validation experiment. Two of the authors manually processed a list of function names in the dataset and the corresponding split from our tokenizer. The task was to evaluate for each split whether it is (1)~perfectly correct, (2)~reasonable, or (3)~wrong. To estimate the accuracy of our tokenizer, we used a subset of 385 randomly selected samples so as to obtain a confidence interval of 95\% with an error margin of 5\%. %
The experiment resulted in 80\% perfect and 95\% at least reasonable splits, averaged between both authors, which confirmed the tokenization to work well in practice. As a reasonable split, we counted
meaning-preserving but slightly off splits such as $\texttt{ndelay\_on} \mapsto \left \{ \texttt{delay}, \texttt{on} \right \}$; wrong splits change the meaning, as in $\texttt{fstarpu\_matrix} \mapsto \left \{ \texttt{fs}, \texttt{tarp}, \texttt{matrix} \right \}$.

\subsection{Label Space}
\label{sec:label-space}

After generating all canonical token sets for the training set we take the union of all string tokens found and count the occurrences of each label. 
The union over all canonical token sets defines our complete label space $L$. However, we define label spaces of varying size to ensure a minimum number of data points per label and explore the impact of an increasing number of labels. 
To this end, \Tool{} generates a new \textit{label space} $L_n$ of size $n$, by
taking the top $n$ most frequently used labels used from the complete label space such that $L_n \subseteq L$.
To obtain the ground truth of labels expressed for each function name, we project each name's canonical set $L_c$ onto those labels that exist in the generated label space $L_n$ as the intersection $L_c \bigcap L_n$.

\subsection{PfastreXML for Function Labels}
\label{sec:label-prediction}

The assignment of function labels, i.e., function name tokens, to functions has many parallels to XML problems such as assigning categories to Wikipedia articles:
compared to the total size of the label space, each single function name has only very few positive labels. The frequency distribution of labels is extremely skewed (see \autoref{fig:dataset-sample-analysis}), as labels like \verb+get+ or \verb+set+ are extremely common (over 36,000 and 20,000 occurrences, respectively), while there is a long tail of infrequent labels ($L_{4096}$ has 2,364 labels with fewer than 100 samples). The re-ranking mechanism of PfastreXML specifically ensures that such infrequent labels do not get lost in the tree-based hierarchy of FastXML.

In addition, there may be more labels that would have been suitable for a given function than just the ones that part of its name in the ground truth. There are often alternative but equivalent ways to name a function because of synonyms among tokens such as \verb+delete+ and \verb+erase+. As a result, we see ``noisy'' data where relevant labels are missing from the ground truth. The propensity-scored loss function in PfastreXML can model the probability of labels to be missing, with appropriate hyper-parameter training.

To remove popularity bias and assign rare labels we weight each label inversely to its popularity. 
\Tool{} optimizes hyper-parameters $A$ and $B$ for the propensity calculation (\autoref{eq:pfastrexml-propensities}) such that the log distribution of propensities matches the sigmoid function. 
Following the recommendation of the PfastreXML authors, we perform a grid search on hyper-parameters $\alpha$ and $\gamma$ close to known good parameter values that optimizes the nDCG calculated on our models validation dataset when trained on a training dataset as discussed in~\autoref{sec:evaluation-embeddings}.

The PfastreXML model predicts the probability distribution of labels for each embedding of a binary function, which induces a ranking over the labels.
To produce a concrete set of predicted labels for multi-label classification and synthesizing a function name, we define a threshold $p_t$ such that \Tool{} outputs those labels with probability greater than $p_t$. 

\section{Function Name Generation}
\label{sec:ngrams-model}

The subset of labels chosen by thresholding is ordered by relevance. We can either present this ranked list of labels directly to the reverse engineer, or we can synthesize a plausible function name containing the labels. Although generating an actual function name loses the ranking information, it may be more easily understood by reverse engineers and can be directly integrated into the reverse engineering process. We now introduce our approach to generating function names~(\autoref{sec:lang-model}) and discuss its accuracy~(\autoref{sec:lang-model-accuracy}).

\subsection{Language Model}
\label{sec:lang-model}

We use a language model for synthesizing plausible function name strings. Our hypothesis is that developers order words in function names following certain rules, defining a form of language where function names are sentences made up of labels.
We face the following problem: given a set of labels, which ordering of the labels would be the most likely in a real function name?  To solve this, we train a language model that can compute a probability score for a given sequence of labels and then pick the order with the highest probability. 

The corpus for training our model consists of all function names split into their constituent labels. We train a language model with modified Kneser-Ney smoothing~\cite{kneser-ney, modified-kneser-ney} that interpolates trigram frequencies over function names with bigrams and unigrams to account for unseen trigrams at test time. We use an efficient implementation by Heafield~\cite{kenlm-queries}, which trains a model from 330k function names in under two seconds on a laptop.

Finding the ordering of labels with maximum probability is essentially the NP-hard maximum Hamiltonian path problem. We use a version of branch and bound to rule out unlikely combinations of labels early and compute scores of orderings using the language model. As a seed solution, we use a simple greedy strategy that picks the most likely trigrams from left to right. Branch and bound is guaranteed to eventually find the optimal ordering, but we use a timeout of 1 million steps to terminate long-running queries with many labels, after which we accept the currently highest-scoring ordering.

\subsection{Accuracy}
\label{sec:lang-model-accuracy}

Using a training/testing split of 9:1 over 330k unique function names, the language model predicts the correct order of labels for 70.0\% of the function names (averaged over 10 runs). On average, each function name is processed in under 2ms. This shows that indeed a classical language model is able to predict useful label orders in most cases, including seemingly hard names such as \texttt{tp\-\_svc\-\_channel\-\_type\-\_streamed\-\_media\-\_emit\-\_stream\-\_state\-\_changed} (from \verb+telepathy-glib+).

The cases where it fails to produce the right order roughly correspond to three groups:
First, often the model produces an order that is simply an alternative to the original one with the same meaning. For instance, it predicts \verb+init_tree+ where \verb+tree_init+ would have been correct, or \verb+alloc_obj+ for \verb+obj_alloc+. 
Second, there are cases with multiple equally plausible orderings with different semantics, e.g., \verb+is_type_array+ vs. \verb+is_array_type+ or \verb+index_to_dir_list+ vs. \verb+dir_list_to_index+.
Third, some labels are just too rare for the language model to meaningfully generalize. For instance, \verb+evict_user_connection+ is ordered as \verb+evict_connection_user+ because the label \textit{evict} occurred only once in that testing split, as part of \verb+evict_connection+.

\section{Evaluation}
\label{sec:evaluation}

We now present our evaluation of \Tool{} and \Embeddings{}. We begin by explaining the makeup of our dataset~(\autoref{sec:evaluation-dataset}) and the computational resources used~(\autoref{sec:evaluation-cost}).
The main evaluation is split into two parts: \autoref{sec:evaluation-embeddings} focuses on comparing different function embeddings on a given task, and \autoref{sec:evaluation-labeling} focuses on comparing different tools for end-to-end function name prediction. This corresponds to two research questions that we answer in the evaluation:

\begin{quote}
	\textbf{RQ1:} \emph{Which binary function embedding is most suited for the task of ranking function labels?}
\end{quote}
To answer this question, we compare \Embeddings{} against the Asm2Vec~\cite{asm2vec}, SAFE~\cite{safe}, and Palmtree~\cite{palmtree} embeddings on the task of function labeling using \Tool{} and demonstrate that \Embeddings{} outperforms the state of the art on this task (\autoref{sec:evaluation-embeddings}).

\begin{quote}
	\textbf{RQ2:} \emph{Does \Tool{} generate more suitable function labels than state-of-the-art approaches?} 
\end{quote}
To answer this question, we compare \Tool{} (using \Embeddings{} embeddings) against the state of the art in function name prediction~(\autoref{sec:evaluation-labeling}). We compare tools both in a ranking task with information-theoretic metrics (\autoref{sec:eval-measure-rank}) and using traditional metrics for multi-class classification (\autoref{sec:eval-ml-classification}). This demonstrates that \Tool{} picks out the most relevant labels for functions and is therefore able to generate the most accurate names. We also specifically investigate the problem of dealing with names a model has never seen before (\autoref{sec:eval-generalization}).

\Embeddings{}, \Tool{}, and the language model are implemented in Python and TensorFlow in about 30 KLOC. Our models and data are available on GitHub\footnote{https://github.com/unibw-patch/xfl}.

\subsection{Dataset}
\label{sec:evaluation-dataset}

Training and testing require a set of ELF binaries with ground truth symbol information that is sufficiently large for generalizing semantics of binary functions for each label. We use the Punstrip dataset~\cite{acsac20-punstrip}, which contains 741,724 functions from 10,047 C binaries taken from pre-compiled Debian packages. These binaries have been compiled with a mixture of compilers and compiler versions from individual package maintainers. 

We use the global symbol bindings in the ELF symbol table as ground truth for obtaining function boundaries and the corresponding names. We exclude pseudo functions of size zero, overlapping functions, and locally bound symbols, which do not clearly correspond to a well-defined function.
Note that in training, the symbol table is available for reading function boundaries. When predicting labels in an unknown binary, symbols have been stripped and function boundaries would have to be obtained using function boundary prediction, for which a number of mature tools exist~\cite{byteweight,shin2015,nucleus}. \Tool{} supports reading function boundaries from the disassemblers Ghidra and Radare2, and the academic tool Nucleus~\cite{nucleus}. 
However, to factor out the performance of function boundary prediction from our evaluation, we equip all tools with function boundaries from the symbol table also for test data.

The hyper-parameters $A$ and $B$ of PfastreXML for computing label propensities (see \autoref{eq:pfastrexml-propensities}) are dataset specific. We compute them for the dataset such that the log distribution of propensities matches the sigmoid function and obtain $A=0.5$ and $B=0.425$ for the Debian dataset.

\begin{table}[t]
    \centering
    \caption{Comparison of the embeddings and the datasets generated for training XFL. Differences are due to limitations in pre-processing. Average labels per point and points per label were calculated for a label space of size 4096.} 
    \label{tab:dataset-analysis}
    \begin{tabularx}{\linewidth}{ Xrrrrr }
        \toprule
        \textbf{Property} & 
        \textbf{\Embeddings{}} &
        \textbf{Asm2Vec} &
        \textbf{PalmTree} &
        \textbf{SAFE} \\
        \midrule
        Model Params.       & 377.0 M   & -   & 3.2 M & 57.2 M \\
        Size                & 512       & 50        & 128    & 100 \\
        Model Type & \hspace*{-3pt}Autoenc. & PV-DM & BERT & Self-Att. \\
		\midrule
		Train Samples       & 400,357   & 396,796   & 386,205 & 342,610 \\
        Test Samples        & 22,422    & 22,224    & 21,405    & 19,035 \\
        Avg. Labels per Point  &   2.93  & 2.93  & 2.85    & 2.75 \\
        Avg. Points per Label  &   320.45  & 317.44  & 298.83  & 255.24 \\
        \bottomrule
    \end{tabularx}
\end{table}

\subsection{Computational Cost}
\label{sec:evaluation-cost}

All of our experiments were carried out on a machine with an AMD EPYC 2 64-Core CPU, a single 16 GB NVIDIA Tesla T4 GPU, and 1 TB of RAM. 

All embedding approaches rely on pre-processing a given binary using disassemblers like Radare2 or IDA Pro, which is required for practically any form of reverse engineering. Training the embeddings is then the step that is most taxing on GPUs. Once a pretrained embedding is available, producing an embedding from an already preprocessed binary is very fast.

For the evaluation of \Embeddings{} and \Tool{}, we cache our binary analysis results and experimentation configurations 
using PostgreSQL and Redis databases. 
We parallelized the binary analysis and were able to finish the preprocessing and feature extraction for the full Debian dataset in 120 hours. %

Training the \Embeddings{} embeddings used about 850 GB of main memory and took 72 hours. %
Once complete, generating embeddings for a pre-extracted feature vector takes an average of 10ms per function. 
Training the Asm2Vec embeddings took about 30 hours.
For SAFE and PalmTree, we used pretrained embeddings so we cannot directly compare cost.

Complete training and testing for an XFL model took 13--15 minutes and 73--85 GB of memory depending on the size of the embedding used, with \Embeddings{} taking most and Asm2Vec least time and memory. \autoref{tab:dataset-analysis} compares the embeddings by their embedding sizes, their model parameters, and their type.

\subsection{Comparing \Embeddings{} against SotA Binary Embeddings}
\label{sec:evaluation-embeddings}

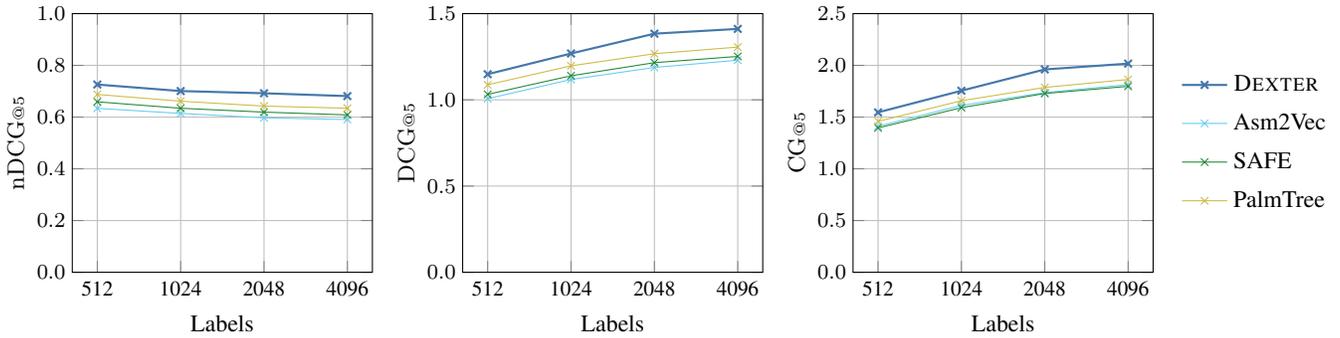
\begin{figure*}[t]
\begin{tikzpicture}
	\pgfplotsset{/tikz/mark=x}
	\begin{groupplot}[
		width=.22\textwidth, 
		scale only axis, 
		group style={
			group size=3 by 1, 
			ylabels at=edge left, 
			horizontal sep=12mm
			},
		xlabel=Labels,
		cycle list={csblue, cscyan, csgreen, csyellow, csred, cspurple, csgrey},
		xmode=log,
		log basis x ={2},
		xtick={512, 1024, 2048, 4096},
		xticklabels={512, 1024, 2048, 4096},
		xmajorgrids,
		ymajorgrids,
		ymin=0,
		ylabel shift=-2pt,
		ylabel style={font=\small},
		xlabel style={font=\small},
		yticklabel shift=-1pt,
		yticklabel style={
			font=\footnotesize,
			/pgf/number format/precision=1,
			/pgf/number format/fixed,
			/pgf/number format/fixed zerofill,
			},
		xticklabel style={font=\footnotesize},
		legend style={font=\small},
		legend cell align={left},
		]
		\nextgroupplot[			
			legend to name=leg_emb, 
			legend columns=1,
			legend style={
				draw=none, 
				row sep=3pt,
			},
			ylabel=$\ndcgat{5}$, 
			ytick distance=0.2,
			ymax=1, 
		]
		\addplot [color=csblue, thick] table [x=l, y=dexter]{\tablendcg};\addlegendentry{\Embeddings};
		\addplot table [x=l, y=asm2vec]{\tablendcg};\addlegendentry{Asm2Vec};
		\addplot table [x=l, y=safe]{\tablendcg};\addlegendentry{SAFE};
		\addplot table [x=l, y=palmtree_gemini]{\tablendcg};\addlegendentry{PalmTree};
		\coordinate (topleft) at (rel axis cs:0,1);
		\coordinate (bottomleft) at (rel axis cs:0,0);

		\nextgroupplot[ylabel=$\dcgat{5}$, ymax=1.5, ytick distance=0.5]
		\addplot [color=csblue, thick] table [x=l, y=dexter]{\tabledcg};
		\addplot table [x=l, y=asm2vec]{\tabledcg};
		\addplot table [x=l, y=safe]{\tabledcg};
		\addplot table [x=l, y=palmtree_gemini]{\tabledcg};

		\nextgroupplot[ylabel=$\cgat{5}$, ymax=2.5, ytick distance=0.5]
		\addplot [color=csblue, thick] table [x=l, y=dexter]{\tablecg};
		\addplot table [x=l, y=asm2vec]{\tablecg};
		\addplot table [x=l, y=safe]{\tablecg};
		\addplot table [x=l, y=palmtree_gemini]{\tablecg};
		\coordinate (topright) at (rel axis cs:1,1);
		\coordinate (bottomright) at (rel axis cs:1,0);
	\end{groupplot}
	\path (topright)--(bottomright) coordinate[midway] (group center);
	\node[right=4pt] at(group center) {\pgfplotslegendfromname{leg_emb}};

\end{tikzpicture}
    \caption{A comparison of the normalized discounted cumulative gain ($\ndcgat{5}$), the discounted cumulative gain ($\dcgat{5}$), and the cumulative gain ($\cgat{5}$) achieved by \Tool{} between \Embeddings{}, Asm2Vec, SAFE, and PalmTree.}
\label{fig:embeddings-results}
\end{figure*}

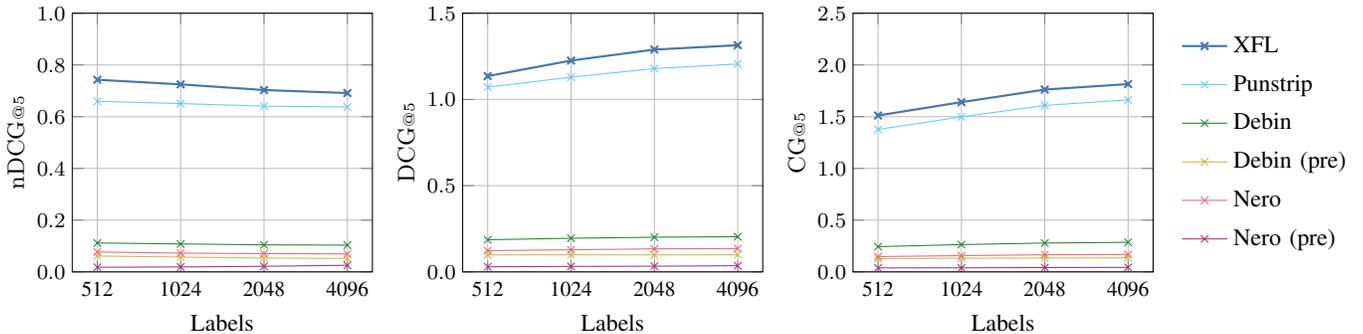
\begin{figure*}[t]
\begin{tikzpicture}
	\pgfplotsset{/tikz/mark=x}
	\pgfkeys{/pgf/number format/.cd,fixed} 
	\begin{groupplot}[
		width=.22\textwidth, 
		scale only axis, 
		group style={
			group size=3 by 1, 
			ylabels at=edge left, 
			horizontal sep=12mm
			},
		xlabel=Labels,
		cycle list={csblue, cscyan, csgreen, csyellow, csred, cspurple, csgrey},
		xmode=log,
		log basis x ={2},
		xtick={512, 1024, 2048, 4096},
		xticklabels={512, 1024, 2048, 4096},
		xmajorgrids,
		ymajorgrids,
		ymin=0,
		ylabel shift=-2pt,
		ylabel style={font=\small},
		xlabel style={font=\small},
		yticklabel shift=-1pt,
		yticklabel style={
			font=\footnotesize,
			/pgf/number format/precision=1,
			/pgf/number format/fixed,
			/pgf/number format/fixed zerofill,
			},
		xticklabel style={font=\footnotesize},
		legend style={font=\small},
		legend cell align={left},
		]
		\nextgroupplot[			
			legend to name=legend, 
			legend columns=1,
			legend style={
				draw=none, 
				row sep=3pt,
			},
			ylabel=$\ndcgat{5}$, 
			ytick distance=0.2,
			ymax=1, 
		]
		\addplot [color=csblue, thick] table [x=l, y=xfl]{\tablendcgtool};\addlegendentry{\Tool{}};
		\addplot table [x=l, y=punstrip]{\tablendcgtool};\addlegendentry{Punstrip};
		\addplot table [x=l, y=debin]{\tablendcgtool};\addlegendentry{Debin};
		\addplot table [x=l, y=debin-pretrained]{\tablendcgtool};\addlegendentry{Debin (pre)};
		\addplot table [x=l, y=nero]{\tablendcgtool};\addlegendentry{Nero};
		\addplot table [x=l, y=nero-pretrained, mark=o]{\tablendcgtool};\addlegendentry{Nero (pre)};
		\coordinate (topleft) at (rel axis cs:0,1);
		\coordinate (bottomleft) at (rel axis cs:0,0);
		
		\nextgroupplot[ylabel=$\dcgat{5}$, ymax=1.5, ytick distance=0.5]
		\addplot [color=csblue, thick] table [x=l, y=xfl]{\tabledcgtool};
		\addplot table [x=l, y=punstrip]{\tabledcgtool};
		\addplot table [x=l, y=debin]{\tabledcgtool};
		\addplot table [x=l, y=debin-pretrained]{\tabledcgtool};
		\addplot table [x=l, y=nero]{\tabledcgtool};
		\addplot table [x=l, y=nero-pretrained, mark=o]{\tabledcgtool};

		\nextgroupplot[ylabel=$\cgat{5}$, ymax=2.5, ytick distance=0.5]
		\addplot [color=csblue, thick] table [x=l, y=xfl]{\tablecgtool};
		\addplot table [x=l, y=punstrip]{\tablecgtool};
		\addplot table [x=l, y=debin]{\tablecgtool};
		\addplot table [x=l, y=debin-pretrained]{\tablecgtool};
		\addplot table [x=l, y=nero]{\tablecgtool};
		\addplot table [x=l, y=nero-pretrained]{\tablecgtool};
		\coordinate (topright) at (rel axis cs:1,1);
		\coordinate (bottomright) at (rel axis cs:1,0);
	\end{groupplot}
	\path (topright)--(bottomright) coordinate[midway] (group center);
	\node[right=4pt] at(group center) {\pgfplotslegendfromname{legend}};
\end{tikzpicture}

	\caption{An information theoretic comparison between Debin, \Tool{}, Nero, and Punstrip with increasing label space sizes. All metrics were taken @5 and a default order was added where non-ranking tools predicted fewer than five labels.}
\label{fig:tool-comparison}
\end{figure*}

We evaluate \Embeddings{} against the state-of-the-art embeddings Asm2Vec~\cite{asm2vec}, SAFE~\cite{safe}, and PalmTree~\cite{palmtree}. %
As we are interested in using the embeddings for an XML task, we compare the performance of each embedding evaluated using cumulative gain-based metrics.
SAFE and PalmTree both rely on contrastive learning approaches that require the same source code to be compiled under different compilation settings. 
As the binaries in our dataset were pre-compiled, we could not use it to create new models and therefore use pretrained models for SAFE\footnote{\url{https://github.com/gadiluna/SAFE}} and PalmTree\footnote{\url{https://github.com/palmtreemodel/PalmTree}} to generate embeddings for the binaries in our dataset. 
For Asm2vec, we used the published code\footnote{\url{https://github.com/McGill-DMaS/Kam1n0-Community}} and applied it to the binaries in our dataset to train a custom model.
To use PalmTree embeddings, we trained a Gemini model to convert the instruction-level embeddings provided into function-level embeddings. With advice from the PalmTree authors, we replicated the setting as described in section 4.4.1 from the original paper~\cite{palmtree} and obtained a model which achieved similar performance\footnote{Our PalmTree Gemini model achieved a testing AUC of 0.939 using a dataset made up from 518 binaries from binutils and coreutils.}. 
We then used \Embeddings{}, Asm2Vec, SAFE, and PalmTree representations of each function to train corresponding XFL models.

Using our dataset we generate embeddings for all functions and randomly split them into a training, validation, and test set using a 90:5:5 ratio. 
After training a PfastreXML model for each approach on the training set, we performed a grid search to fine-tune the model hyper-parameters over our validation dataset, and report the evaluation results on the test set. Using values close to known 
good parameters estimated from the PfastreXML paper, we optimize the $\alpha$ and $\gamma$ hyper-parameters to minimize our nDCG loss (see \autoref{sec:background-pfastrexml}). %
We run experiments with four label spaces $L_n$ with $n \in \{ 512, 1024, 2048, 4096 \}$ to see any relative differences in information gain caused by label space size.

\autoref{fig:embeddings-results} shows the corresponding mean $\ndcgat{5}$, $\dcgat{5}$, and $\cgat{5}$ for all embeddings.
We can see that \Embeddings{} outperforms Asm2Vec, SAFE, and PalmTree embeddings across all label space sizes. Among the other three, PalmTree clearly dominates; in their paper, Li et al.~\cite{palmtree} demonstrate that the semantics-related components are important factors. Using pre-processing and manual feature engineering \Embeddings{} includes even more code semantics and thus can outperform pure representation learning on the binary code domain.

\subsection{Comparing \Tool{} against SotA in Function Naming}
\label{sec:evaluation-labeling}

We evaluate \Tool{} against the three state-of-the-art function name prediction tools Debin~\cite{debin}, Nero~\cite{nero}, and Punstrip~\cite{acsac20-punstrip}. 
We also include a comparison to published results of a customized version of Dire~\cite{dire} taken from the Nero paper~\cite{nero} (the Dire authors recommended us not to use their tool for function name prediction).
Nero combines heavy static analysis to obtain an Augmented Control Flow Graph and feeds this representation into state-of-the-art neural networks. In their work, a variant based on a Graph Neural Network (GNN) performs best, so we compare against this. 
Debin~\cite{debin} and Punstrip~\cite{acsac20-punstrip} both make predictions for function names based on Conditional Random Fields to compute a maximal joint probability of function name assignments in a binary.
For Debin and Nero, we compare against both a pretrained model released by the authors and a custom model trained on our dataset. For Punstrip, no pretrained model is available. %

We first randomly split our dataset obtained in~\autoref{sec:evaluation-dataset} into training, validation, and testing sets with a respective 90:5:5 ratio. 
For each of the tools, we train a model on binaries in the training set, tune model hyper-parameters on the validation set, and show our results against the model's predictions on the test dataset. The list of binaries in all three sets are kept consistent when evaluating all of the tools. We use the same four label space sizes as before.

We evaluate all tools on two problems, both related to predicting relevant labels contained in function names. 
The first problem~(\autoref{sec:eval-measure-rank}) evaluates the measure of rank, whereby we produce a ranked list of labels in the label space for each data point. We assign each correct/relevant label an equal weight. The second problem~(\autoref{sec:eval-ml-classification}) evaluates each tool in a multi-label classification setting predicting a variably-sized set of relevant labels.
Note that Debin and Punstrip are designed to predict whole function names. Therefore, we use our canonicalization step to project the predicted function names onto our generated label space.
Nero predicts labels itself, but we report here the results of using Nero on our label spaces. We confirmed experimentally that Nero's performance evaluated in this way is in line with the values reported by their own pipeline. %

\subsubsection{Measure of Rank}
\label{sec:eval-measure-rank}
We evaluate this experiment using the information gain metrics $\ndcgat{5}$, $\dcgat{5}$, and $\cgat{5}$ as depicted in~\autoref{fig:tool-comparison}.
As the size of the label space increases, every tool evaluated performed worse in terms of nDCG but better in terms of DCG and CG. With a larger label space, there are more labels per function to be learned. Accordingly, the growing CG and DCG show that more information is learned by the models; still, the normalization against the total number of labels per data point leads to a lower nDCG. 

Our results show that \Tool{} consistently outperforms other state-of-the-art tools in terms of information gain irrespective of label space size; this provides evidence for our hypothesis that predicting labels outperforms predicting whole function names.
Surprisingly, while Nero outperforms Debin on the Nero dataset for function name prediction (\autoref{sec:eval-ml-classification}), it performs worse than Debin on the Debian dataset. We investigated this anomaly to rule out mistakes in our experimental setup, but we were able to reproduce the published values on the Nero dataset, and correspondence with the Nero authors confirmed our results.
We believe this to underline the risks of dataset bias in contemporary approaches to machine learning on binary code, which we attempt to counter with an increased dataset size and an embedding-based approach.

Note that the results for \Tool{} do not match those for \Embeddings{} in \autoref{sec:evaluation-embeddings} due to using different dataset splits: for the embeddings experiment, we randomly split functions independently into training, validation, and testing, whereas in the multi-label classification experiment, we
ensured that functions from the same binary are always in the same set.

\subsubsection{Multi-label Classification}
\label{sec:eval-ml-classification}

\begin{table}
    \centering
    \caption{Multi-label classification evaluation against state-of-the-art tools for the Debian and Nero datasets, highest values in bold. Results marked with $\dagger$ are taken directly from \cite{nero} without us rerunning the experiment.} 
    \label{tab:multilabel-eval-results}
	\newcommand{\tabstrut}{\rule{0em}{2.5ex}}
	\newcommand{\zsdag}{{$\mathrlap{^\dagger}$}}
    \begin{tabularx}{\columnwidth}{Xcccccc}
        \toprule
        & \multicolumn{3}{c}{\bf Debian Dataset} & \multicolumn{3}{c}{\bf Nero Dataset} \\
        \midrule
        \textbf{Tool} & 
        \textbf{Prec.} & \textbf{Recall} & $\mathbf{F_1}$ &
        \textbf{Prec.} & \textbf{Recall} & $\mathbf{F_1}$
        \\
        \midrule
        \Tool{}\tabstrut     &  \textbf{0.8345}   &   0.5750  &   \textbf{0.6809}  & \textbf{0.8664} & \textbf{0.4383} & \textbf{0.5821} \\
        Nero\tabstrut        &  0.1600   &   0.0622  &   0.0896  & 0.4861 & 0.4282 & 0.4553 \\
        Debin\tabstrut       &  0.1564   &   0.1081  &   0.1279  & 0.3486\zsdag & 0.3254\zsdag & 0.3366\zsdag \\
        Punstrip\tabstrut    &  0.6336   &   \textbf{0.6350}  &  0.6343  & -     & -     & - \\
        Dire\tabstrut        &  -        &   -       &   -       & 0.3802\zsdag   & 0.3333\zsdag     & 0.3552\zsdag \\
        \bottomrule
    \end{tabularx}
\end{table}

To evaluate \Tool{}'s performance in a multi-label classification task on standard metrics, we modify the experiment used in the measure of rank setup to 
include a linear threshold $p_t$ to decide if the label is relevant. This results in a variably sized subset of predicted relevant labels for each data point.
We can then define true positives as the number of correct labels predicted with a threshold greater than $p_{t}$. 
False positives are defined by labels which are predicted but not present in the ground truth, and false negatives are correct labels that are missed by our prediction.
During validation, we set $p_t$ to a value that maximizes the F$_1$ score on the validation dataset.

Our results as shown in~\autoref{tab:multilabel-eval-results} display the micro-averaged multi-label precision, recall, and F$_1$ score on both the Debian and Nero datasets using a label space of size 1024. 
\Tool{} outperforms other state-of-the-art tools in precision and F$_1$, with Punstrip coming closest and winning on recall with the current $p_t$.
By varying the threshold parameter $p_t$, one would be able to adjust the precision and recall trade off.

We also include an evaluation on the Nero dataset, given the relative difference in performance of Debin and Nero on those binaries.
While \Tool{} achieves comparable, if slightly worse results on this smaller dataset, Debin and Nero both perform significantly better. This shows that \Tool{} is able to benefit from a larger dataset to achieve more generalized representations, whereas Debin and Nero both significantly drop in performance when scaling up.
Apart from the dataset size, another factor may be that in Nero's test set of 13 binaries, many are from the same source repository and are all compiled with the same settings. In contrast, our Debian dataset contains 503 binaries in the test set, which are compiled by individual package maintainers where the compiler, version, and optimization levels are not fixed.

By ordering relevant labels using the language model and joining them by an underscore or in camel case, we can synthesize whole function names. \autoref{tab:function-name-generation} shows complete examples of such names, including typical errors that can be introduced through the function name generation process.
These examples show that our unsupervised tokenization process was automatically able to deduce the importance of the labels \textit{HID} for \textit{human input device}, \textit{lkf} for \textit{locked file}, and \textit{mcx} as a term used in the Markov Clustering set of algorithms.

\begin{table*}[t]
    \caption{Example data points of the function name generation process. Ground truth names are first split into known labels which we aim to predict. The predicted labels are then used to generate a function name according to the language model. Labels from the canonicalization process that did not make it into the label space are crossed out.}
    \label{tab:function-name-generation}
    \renewcommand{\arraystretch}{1.1} %
	\centering
    \begin{tabularx}{.95\textwidth}{Xccl}
        \toprule
        \textbf{Original Name} & 
        \textbf{Ground Truth Tokens} & 
        \textbf{Predicted Labels} & 
        \textbf{Output Name} \\
        
        \midrule

\verb|grub_crypto_cbc_decrypt| & \{grub, cbc, de, crypt\}  & \{de, crypt, grub, crypt, cb, odisk\} & \verb|grub_crypt_odisk_de_cb| \\

        \verb|make_smooth_colormap| &  \{make, color, map, smooth\}     & \{make, color, map, random\}    & \verb|make_random_color_map| \\
\verb|mcxRealloc|           & \{realloc, mcx\}                         & \{realloc, mcx\}                & \verb|mcx_realloc| \\
\verb|check_audio_range|    & \{range, audio, check\}                &   \{audio, range, check\} & \verb|audio_range_check| \\
\verb|HIDSetItemValue|      & \{set, item, hid, value\}        & \{set, item, hid, value, is\} & \verb|is_item_hid_set_value| \\
\verb|mcxTingRelease|       & \{release, mcx, \sout{ting}\}                      & \{release, mcx\}                & \verb|mcx_release| \\
\verb|chirp_recursive_put|  & \{put, recursive, chirp\}         & \{put, recursive, chirp, ticket\}   & \verb|ticket_chirp_recursive_put| \\
\verb|lkfopendata|          & \{lkf, data, open\}                  & \{lkf, open, switches, data\} & \verb|switches_lkf_open_data| \\

\verb|cmdline_parser_init| & \{line, init, cmd, parser\}  &   \{parser, init, cmd, line, csu\} & \verb|csu_cmd_line_parser_init| \\

        \bottomrule
    \end{tabularx}
\end{table*}

\subsubsection{Generalization}
\label{sec:eval-generalization}

Although function \emph{bodies} are never shared between training and testing sets, many function \textit{names} do occur in both. Function names like \verb+hash+, \verb+get_line+, or \verb+usage+ are used frequently by developers, although the corresponding function bodies will be quite different. In practice, we expect such cases to occur frequently, and predicting such names correctly would be useful for a reverse engineer. 

However, multi-label learning for function name generation has the advantage of being able to generalize to new function names that have never been seen before, which is impossible for multi-class approaches. Therefore, we now specifically investigate how well all approaches are able to identify combinations of labels when no function names are shared between the training and testing sets.
To this end, we repeated our experiments as in~\autoref{sec:eval-ml-classification}, but restricted the test set to function names not present in training.

\autoref{tab:generalization} shows that performance drops for every tool but also confirms that \Tool{} is able to predict labels in unseen names better than other approaches. 
Note that while Punstrip and Debin cannot predict unseen function names by design, applying our tokenization to their predictions still yielded some correct labels.
Common labels recovered correctly as part of unknown function names include \verb+get+, \verb+set+, \verb+new+, \verb+free+, and the OCaml-specific tokens \verb+caml+ and \verb+fun+.

\begin{table}
	\centering
	\newcommand{\tabstrut}{\rule{0em}{2.5ex}}
    \caption{Comparison of the ability to predict completely unseen function names (Debian).} 
    \label{tab:generalization}
    \begin{tabularx}{.9\columnwidth}{Xcccccc}
        \toprule
        \textbf{Tool} & \textbf{Prec.} & \textbf{Recall} & $\mathbf{F_1}$ \\
        \midrule
        \Tool{}\tabstrut     &  \textbf{0.4342} & \textbf{0.1739} & \textbf{0.2416} \\
        Nero\tabstrut        &  0.0494 & 0.0238 & 0.0321  \\
        Debin\tabstrut       &  0.0453 & 0.0292 & 0.0380   \\
        Punstrip\tabstrut    &  0.1170 & 0.1190 & 0.1180   \\
        \bottomrule
    \end{tabularx}
\end{table}

\section{Discussion}
\label{sec:discussion}

We now discuss our results and current limitations of our approach, and point out directions for future work.

\subsection{Generalization and Dataset Diversity}
Our results for multi-label classification show that Nero is outperformed by Debin on the Debian dataset; a surprising outcome given that Nero outperforms Debin on the Nero dataset. We contacted the authors of Nero to investigate further and they suggested the following possible causes:
	(i) Nero's dataset limited the maximum size of binaries to 1MB. This may influence the inference of common function names available in most \textit{libc} binaries.
    (ii) The vocabulary size in our dataset is significantly bigger and as a result Nero predicts empty labels $43\%$ of the time.
    (iii) Nero's configuration and implementation had not been fine-tuned to our dataset.
This highlights an interesting issue. Even when taking care to avoid overfitting and ensuring proper train-test splits, there are aspects of the model that are inherently dataset-specific.
We can see that even our own model decreases in performance when testing on the Nero dataset, which uses different compilers and build settings. 

Although our preprocessing and analysis should reduce the amount of data needed, even the Debian dataset is still much smaller than the corpuses used for training state-of-the-art models for natural language processing.
We believe that public, standardized datasets are the way forward, both for source and binary code-based tasks.

\subsection{Application Domains}

Using machine learning for predicting function names inherently requires a training dataset whose distribution resembles the target dataset. That is, we cannot hope to predict function names in Windows binaries with a model trained on GNU/Linux. As it stands, the model we trained for \Tool{} can identify labels in binaries compiled for GNU/Linux systems only, but the target binaries can be closed source. Because of our multi-label approach, \Tool{} can construct suitable names from tokens for function names it has never seen in the training set (see also \autoref{sec:eval-generalization}). 
This can make \Tool{} useful in practice today, for reverse engineering closed-source applications or device drivers, or for forensic analysis of GNU/Linux malware. 
However, as no ground truth is available for closed source software, evaluating this aspect will require a study with human participants. We believe this to be an interesting avenue for future work, in line with recent pioneering work into observational studies on reverse engineers~\cite{VotipkaRMFM20,Mantovani22}.

In principle, a Windows-specific model for \Tool{} could be trained, too, as long as we can construct a sufficiently large ground truth dataset. A dataset the size of the Debian dataset would be challenging to obtain, but not impossible. For instance, Github contains the source code of over 230,000 C files with a \texttt{WinMain} function defined. Another possibility are to use public servers with debug symbols for ground truth, or the HexRays Lumina server, which stores function names for reversed closed-source binaries.
Similarly, our model is specific to the \verb|x86_64| architecture. Because our analysis is built on the VEX IR used by angr, one could use the same tool chain for analyzing other instruction set architectures. 

Our entire toolchain and setup is currently geared towards binaries compiled from C, only. Apart from the differences in distribution that would come with code compiled from other languages, the different naming conventions, namespaces and the resulting name mangling would require changes in our preprocessing infrastructure.

\subsection{Impact of Function Boundaries}
Our evaluation assumes that precise function boundaries are available. At training time, this is a safe assumption as we require debug symbols regardless. At prediction time, tools such as Nucleus~\cite{nucleus} are highly accuracte but can detect more or fewer functions than are actually present. \Tool{} will predict labels for whatever function body it is given, so it would attempt to predict labels for parts of a function or for the combination of multiple functions. When used interactively, a reverse engineer would have to query \Tool{} again for any functions they change boundaries for.

\subsection{Concept Drift}
Even on the same architecture, the testing data distribution can change over time as source code and compilers change. 
This \textit{concept drift} is a known phenomenon in many application domains. The performance differences between different datasets suggest that it is present in the problem of function labeling. It is possible to counteract concept drift through periodic re-training with labeled data. 
Recent work by \citet{CADE} uses contrastive learning to detect the and explain concept drift in new data samples. Such an approach may also be combined with \Tool{} to mitigate the impact of changing test data over time.

\subsection{Adversarial Settings}

In this paper, we rule out an adversarial setting where the developers actively try to prevent recognition of functions by automated means. We explicitly assume the code of functions to be generated by a regular compiler, with only the debug and symbol information to be removed as is typically done for release builds. 
Obfuscation methods like runtime packing or opaque predicates that prevent successful disassembly would break our pipeline without explicit countermeasures~\cite{obfuscation-csur16}. Similarly, more subtle obfuscation methods would substantially change the character of functions and thus likely sufficiently affect the features to increase mispredictions. 

Beyond obfuscation, one could consider adversarial attacks against the multi-label classifier model in \Tool{}~\cite{HuKWL21,SongJHH18}. Using code transformations as adversarial perturbations~\cite{CastroGPRPC21}, an attacker could intentionally cause mislabeling of functions with specific labels. Adversarial robustness is an active area of research, and most proposed countermeasures have quickly been shown to be ineffective~\cite{Carlini-robustness19}.

\section{Related Work}
\label{sec:related-work}

We now review related work on the problem of binary function labeling, binary code similarity, and of learning source-level code representations.

\Tool{} predicts labels for functions and is thus related to similar projects which produce labels or names for functions.
Punstrip~\cite{acsac20-punstrip} uses Conditional Random Fields (CRFs) to capture the dependencies between generated fingerprints of functions and their callers and callees. 
Debin~\cite{debin} attempts to recover function names and other debug information such as program variables and their mapping to registers and memory offsets. 
In Debin, CRFs represent relationships between code and data and are used to predict properties of extracted memory cells. 
Nero~\cite{nero} builds an \emph{augmented control flow graph} which extends a CFG with call sites, specifically engineered for procedure name prediction. The representation is fed into GNNs, LSTMs and Transformer architectures.
Dire~\cite{dire} only supports variable name prediction; however, the authors of Nero modified the project to support 
the prediction of procedure names. Dire uses an encoder-decoder neural network, taking as input both tokenized code and the AST from a decompiler and generates embeddings for each identifier which are used by the decoder to predict names.

Much recent work on vector embeddings for binary code focuses on the task of \emph{binary code similarity}. DeepBinDiff~\cite{duan2020deepbindiff} defines the task as trying to find the best match between similar basic blocks based upon their control flow dependency. A different approach is to use the \emph{same source policy} which defines two binaries or functions to be similar if they are compiled from the same source code but for different target architectures~\cite{gemini}, different source code versions~\cite{alphadiff} or different compilers and compiler settings~\cite{asm2vec}. 

Many approaches borrow from natural language processing. 
We already discussed SAFE~\cite{safe}, Asm2Vec~\cite{asm2vec}, and PalmTree~\cite{palmtree} in \autoref{sec:embeddings-soa}. 
Zuo et. al~\cite{zuo2019neural} rely on Word2Vec, but adapt \emph{Neural Machine Translation} to handle single instructions as words and basic blocks as sentences.
Sun et al.~\cite{sun2019binary}
use approaches found in bioinformatics such as longest common sub-sequence algorithms in conjunction with Word2Vec embeddings to measure binary semantic similarity.

While \Tool{} uses debug information from the compilation process as ground truth, our model relies only on information found in stripped binaries, i.e. \emph{without} access to the source code. Nevertheless, we review source code based approaches to function similarity.
Source code function similarity has been used in program comprehension, function name suggestion, and source code completion. Models for source code utilize syntax information from Abstract Syntax Trees~\cite{phog}~\cite{alon18}.
Program representations may be constructed using generative models~\cite{brockschmidt2018generative}, graph neural networks~\cite{allamanis2018learning},  graph models enriched with sequence encoders~\cite{fernandes2018structured}, or attention-based models~\cite{alon2018codeseq}. A convolutional network is used to summarize source code to tokens~\cite{allamanis2016convolutional}.

Another related area is code authorship attribution, i.e., identifying, verifying, or clustering code authors; we refer to \citet{kalgutkar2019code} for a recent survey.

\section{Conclusion}
\label{sec:conclusion}

We present \Embeddings{} and \Tool{} to solve the function naming problem, addressing limitations in earlier methods.

\Embeddings{} creates a distributed representation of binary code that concisely captures the semantics of functions. 
Our embeddings condense millions of features drawn from the whole binary, the function's calling context, and the function itself. %
We show that it outperforms state-of-the-art binary code embeddings when used for predicting labels in function names. This provides evidence that using static analysis results for learning embeddings improves performance in comparison to relying on just learning on raw syntax. 

\Tool{} uses \Embeddings{} to perform multi-label classification and learn an XML model to predict common tokens found in the names of functions from C binaries in Debian. 
We show that our approach outperforms existing approaches to function name prediction.
In particular, \Tool{} is able to predict names for functions even when no function of that name is contained in the training set.

\bibliographystyle{IEEEtranN}

\begin{thebibliography}{69}
\providecommand{\natexlab}[1]{#1}
\providecommand{\url}[1]{#1}
\csname url@samestyle\endcsname
\providecommand{\newblock}{\relax}
\providecommand{\bibinfo}[2]{#2}
\providecommand{\BIBentrySTDinterwordspacing}{\spaceskip=0pt\relax}
\providecommand{\BIBentryALTinterwordstretchfactor}{4}
\providecommand{\BIBentryALTinterwordspacing}{\spaceskip=\fontdimen2\font plus
\BIBentryALTinterwordstretchfactor\fontdimen3\font minus
  \fontdimen4\font\relax}
\providecommand{\BIBforeignlanguage}[2]{{%
\expandafter\ifx\csname l@#1\endcsname\relax
\typeout{** WARNING: IEEEtranN.bst: No hyphenation pattern has been}%
\typeout{** loaded for the language `#1'. Using the pattern for}%
\typeout{** the default language instead.}%
\else
\language=\csname l@#1\endcsname
\fi
#2}}
\providecommand{\BIBdecl}{\relax}
\BIBdecl

\bibitem[Chikofsky and II(1990)]{ChikofskyC90}
E.~J. Chikofsky and J.~H.~C. II, ``Reverse engineering and design recovery: {A}
  taxonomy,'' \emph{{IEEE} Softw.}, vol.~7, no.~1, pp. 13--17, 1990.

\bibitem[Arce(2002)]{bughunting02}
I.~Arce, ``Bug hunting: The seven ways of the security samurai (supplement to
  computer magazine),'' \emph{IEEE Computer}, vol.~35, no.~04, pp. 11--15,
  2002.

\bibitem[Cifuentes(1999)]{Cifuentes99}
C.~Cifuentes, ``The impact of copyright on the development of cutting edge
  binary reverse engineering technology,'' in \emph{Working Conf. Reverse
  Engineering (WCRE)}.\hskip 1em plus 0.5em minus 0.4em\relax {IEEE} Computer
  Society, 1999, pp. 66--76.

\bibitem[Siksorski and Honig(2012)]{sikorski-practicalmalware}
M.~Siksorski and A.~Honig, \emph{Practical Malware Analysis: The Hands-On Guide
  to Dissecting Malicious Software}.\hskip 1em plus 0.5em minus 0.4em\relax San
  Francisco: No Starch Press, 2012.

\bibitem[Votipka et~al.(2020)Votipka, Rabin, Micinski, Foster, and
  Mazurek]{VotipkaRMFM20}
D.~Votipka, S.~M. Rabin, K.~K. Micinski, J.~S. Foster, and M.~L. Mazurek, ``An
  observational investigation of reverse engineers' processes,'' in \emph{29th
  {USENIX} Security Symposium, {USENIX} Security 2020}.\hskip 1em plus 0.5em
  minus 0.4em\relax {USENIX} Association, 2020, pp. 1875--1892.

\bibitem[Mantovani et~al.(2022)Mantovani, Aonzo, Fratantonio, and
  Balzarotti]{Mantovani22}
A.~Mantovani, S.~Aonzo, Y.~Fratantonio, and D.~Balzarotti, ``Re-mind: a first
  look inside the mind of a reverse engineer,'' in \emph{Proc. 31st {USENIX}
  Security Symposium (USENIX Security)}.\hskip 1em plus 0.5em minus 0.4em\relax
  {USENIX} Association, 2022.

\bibitem[Bichsel et~al.(2016)Bichsel, Raychev, Tsankov, and Vechev]{debin}
B.~Bichsel, V.~Raychev, P.~Tsankov, and M.~T. Vechev, ``Statistical
  deobfuscation of android applications,'' in \emph{{ACM} {SIGSAC} Conference
  on Computer and Communications Security (CCS)}.\hskip 1em plus 0.5em minus
  0.4em\relax {ACM}, 2016, pp. 343--355.

\bibitem[Patrick-Evans et~al.(2020)Patrick-Evans, Cavallaro, and
  Kinder]{acsac20-punstrip}
J.~Patrick-Evans, L.~Cavallaro, and J.~Kinder, ``Probabilistic naming of
  functions in stripped binaries,'' in \emph{Proc. 35th Annu. Computer Security
  Applications Conference (ACSAC)}.\hskip 1em plus 0.5em minus 0.4em\relax ACM,
  2020, pp. 373--385.

\bibitem[Joachims(1998)]{Joachims98}
T.~Joachims, ``Text categorization with support vector machines: Learning with
  many relevant features,'' in \emph{European Conf. Machine Learning (ECML)},
  ser. LNCS, vol. 1398.\hskip 1em plus 0.5em minus 0.4em\relax Springer, 1998,
  pp. 137--142.

\bibitem[Schapire and Singer(2000)]{SchapireS00}
R.~E. Schapire and Y.~Singer, ``{BoosTexter}: {A} boosting-based system for
  text categorization,'' \emph{Mach. Learn.}, vol.~39, no. 2/3, pp. 135--168,
  2000.

\bibitem[Nigam et~al.(2000)Nigam, McCallum, Thrun, and Mitchell]{NigamMTM00}
K.~Nigam, A.~McCallum, S.~Thrun, and T.~M. Mitchell, ``Text classification from
  labeled and unlabeled documents using {EM},'' \emph{Mach. Learn.}, vol.~39,
  no. 2/3, pp. 103--134, 2000.

\bibitem[Prabhu and Varma(2014)]{PrabhuV14}
Y.~Prabhu and M.~Varma, ``{FastXML}: a fast, accurate and stable
  tree-classifier for extreme multi-label learning,'' in \emph{The 20th {ACM}
  {SIGKDD} International Conference on Knowledge Discovery and Data Mining,
  {KDD}}.\hskip 1em plus 0.5em minus 0.4em\relax {ACM}, 2014, pp. 263--272.

\bibitem[Bhatia et~al.(2015)Bhatia, Jain, Kar, Varma, and Jain]{BhatiaJKVJ15}
K.~Bhatia, H.~Jain, P.~Kar, M.~Varma, and P.~Jain, ``Sparse local embeddings
  for extreme multi-label classification,'' in \emph{Annu. Conf. Neural
  Information Processing Systems (NIPS)}, 2015, pp. 730--738.

\bibitem[Li et~al.(2021)Li, Qu, and Yin]{palmtree}
X.~Li, Y.~Qu, and H.~Yin, ``{PalmTree}: Learning an assembly language model for
  instruction embedding,'' in \emph{Proc. {ACM} {SIGSAC} Conf. Computer and
  Communications Security (CCS)}.\hskip 1em plus 0.5em minus 0.4em\relax {ACM},
  2021, pp. 3236--3251.

\bibitem[Massarelli et~al.(2019)Massarelli, Luna, Petroni, Baldoni, and
  Querzoni]{safe}
L.~Massarelli, G.~A.~D. Luna, F.~Petroni, R.~Baldoni, and L.~Querzoni,
  ``{SAFE:} self-attentive function embeddings for binary similarity,'' in
  \emph{Detection of Intrusions and Malware, and Vulnerability Assessment},
  vol. 11543.\hskip 1em plus 0.5em minus 0.4em\relax Springer, 2019, pp.
  309--329.

\bibitem[Ding et~al.(2019)Ding, Fung, and Charland]{asm2vec}
S.~H.~H. Ding, B.~C.~M. Fung, and P.~Charland, ``Asm2vec: Boosting static
  representation robustness for binary clone search against code obfuscation
  and compiler optimization,'' in \emph{{IEEE} Symposium on Security and
  Privacy}.\hskip 1em plus 0.5em minus 0.4em\relax {IEEE}, 2019, pp. 472--489.

\bibitem[Mukherjee et~al.(2021)Mukherjee, Wen, Chaudhari, Reps, Chaudhuri, and
  Jermaine]{MukherjeeWCRCJ21}
R.~Mukherjee, Y.~Wen, D.~Chaudhari, T.~W. Reps, S.~Chaudhuri, and C.~M.
  Jermaine, ``Neural program generation modulo static analysis,'' in
  \emph{Annu. Conf. Neural Information Processing Systems (NeurIPS)}, 2021, pp.
  18\,984--18\,996.

\bibitem[Allwein et~al.(2000)Allwein, Schapire, and Singer]{AllweinSS00}
E.~L. Allwein, R.~E. Schapire, and Y.~Singer, ``Reducing multiclass to binary:
  {A} unifying approach for margin classifiers,'' \emph{J. Mach. Learn. Res.},
  vol.~1, pp. 113--141, 2000.

\bibitem[Tai and Lin(2012)]{TaiL12}
F.~Tai and H.~Lin, ``Multilabel classification with principal label space
  transformation,'' \emph{Neural Comput.}, vol.~24, no.~9, pp. 2508--2542,
  2012.

\bibitem[Ciss{\'{e}} et~al.(2013)Ciss{\'{e}}, Usunier, Arti{\`{e}}res, and
  Gallinari]{CisseUAG13}
M.~Ciss{\'{e}}, N.~Usunier, T.~Arti{\`{e}}res, and P.~Gallinari, ``Robust bloom
  filters for large multilabel classification tasks,'' in \emph{Advances in
  Neural Information Processing Systems 26}, 2013, pp. 1851--1859.

\bibitem[Yu et~al.(2014)Yu, Jain, Kar, and Dhillon]{YuJKD14}
H.~Yu, P.~Jain, P.~Kar, and I.~S. Dhillon, ``Large-scale multi-label learning
  with missing labels,'' in \emph{Proc. 31st Int. Conf. Machine Learning
  (ICML)}, ser. {JMLR} Workshop and Conference Proceedings, vol.~32, 2014, pp.
  593--601.

\bibitem[Dahiya et~al.(2021)Dahiya, Saini, Mittal, Shaw, Dave, Soni, Jain,
  Agarwal, and Varma]{Dahiya21}
K.~Dahiya, D.~Saini, A.~Mittal, A.~Shaw, K.~Dave, A.~Soni, H.~Jain, S.~Agarwal,
  and M.~Varma, ``Deepxml: A deep extreme multi-label learning framework
  applied to short text documents,'' in \emph{Proceedings of the ACM
  International Conference on Web Search and Data Mining}, March 2021.

\bibitem[Jain et~al.(2019)Jain, Balasubramanian, Chunduri, and Varma]{slice}
H.~Jain, V.~Balasubramanian, B.~Chunduri, and M.~Varma, ``Slice: Scalable
  linear extreme classifiers trained on 100 million labels for related
  searches,'' in \emph{Proc. ACM Int. Conf. Web Search and Data Mining
  (WSDM)}.\hskip 1em plus 0.5em minus 0.4em\relax {ACM}, 2019, pp. 528--536.

\bibitem[Mittal et~al.(2021)Mittal, Dahiya, Agrawal, Saini, Agarwal, Kar, and
  Varma]{Mittal21}
A.~Mittal, K.~Dahiya, S.~Agrawal, D.~Saini, S.~Agarwal, P.~Kar, and M.~Varma,
  ``Decaf: Deep extreme classification with label features,'' in
  \emph{Proceedings of the ACM International Conference on Web Search and Data
  Mining}, March 2021.

\bibitem[Bengio et~al.(2010)Bengio, Weston, and Grangier]{BengioWG10}
S.~Bengio, J.~Weston, and D.~Grangier, ``Label embedding trees for large
  multi-class tasks,'' in \emph{Annu. Conf. Neural Information Processing
  Systems (NeurIPS)}.\hskip 1em plus 0.5em minus 0.4em\relax Curran Associates,
  Inc., 2010, pp. 163--171.

\bibitem[Deng et~al.(2011)Deng, Satheesh, Berg, and Fei{-}Fei]{DengSBL11}
J.~Deng, S.~Satheesh, A.~C. Berg, and L.~Fei{-}Fei, ``Fast and balanced:
  Efficient label tree learning for large scale object recognition,'' in
  \emph{Annu. Conf. Neural Information Processing Systems (NIPS)}, 2011, pp.
  567--575.

\bibitem[Agrawal et~al.(2013)Agrawal, Gupta, Prabhu, and Varma]{AgrawalGPV13}
R.~Agrawal, A.~Gupta, Y.~Prabhu, and M.~Varma, ``Multi-label learning with
  millions of labels: recommending advertiser bid phrases for web pages,'' in
  \emph{Proc. World Wide Web Conf. (WWW)}.\hskip 1em plus 0.5em minus
  0.4em\relax ACM, 2013, pp. 13--24.

\bibitem[Weston et~al.(2013)Weston, Makadia, and Yee]{WestonMY13}
J.~Weston, A.~Makadia, and H.~Yee, ``Label partitioning for sublinear
  ranking,'' in \emph{Proc. 30th Int. Conf. Machine Learning (ICML)}, ser.
  {JMLR} Workshop and Conference Proceedings, vol.~28.\hskip 1em plus 0.5em
  minus 0.4em\relax JMLR.org, 2013, pp. 181--189.

\bibitem[Balasubramanian and Lebanon(2012)]{BalasubramanianL12}
K.~Balasubramanian and G.~Lebanon, ``The landmark selection method for multiple
  output prediction,'' in \emph{Proc. 29th Int. Conf. Machine Learning
  (ICML)}.\hskip 1em plus 0.5em minus 0.4em\relax icml.cc / Omnipress, 2012.

\bibitem[J{\"{a}}rvelin and Kek{\"{a}}l{\"{a}}inen(2002)]{JarvelinK02}
K.~J{\"{a}}rvelin and J.~Kek{\"{a}}l{\"{a}}inen, ``Cumulated gain-based
  evaluation of {IR} techniques,'' \emph{{ACM} Trans. Inf. Syst.}, vol.~20,
  no.~4, pp. 422--446, 2002.

\bibitem[Hsu et~al.(2009)Hsu, Kakade, Langford, and Zhang]{HsuKLZ09}
D.~J. Hsu, S.~M. Kakade, J.~Langford, and T.~Zhang, ``Multi-label prediction
  via compressed sensing,'' in \emph{Annu. Conf. Neural Information Processing
  Systems (NeurIPS)}.\hskip 1em plus 0.5em minus 0.4em\relax Curran Associates,
  Inc., 2009, pp. 772--780.

\bibitem[Weston et~al.(2011)Weston, Bengio, and Usunier]{WestonBU11}
J.~Weston, S.~Bengio, and N.~Usunier, ``{WSABIE:} scaling up to large
  vocabulary image annotation,'' in \emph{Proc. Int. Joint Conf. on Artificial
  Intelligence (IJCAI)}.\hskip 1em plus 0.5em minus 0.4em\relax {IJCAI/AAAI},
  2011, pp. 2764--2770.

\bibitem[Jain et~al.(2016)Jain, Prabhu, and Varma]{JainPV16}
H.~Jain, Y.~Prabhu, and M.~Varma, ``Extreme multi-label loss functions for
  recommendation, tagging, ranking {\&} other missing label applications,'' in
  \emph{{SIGKDD} International Conference on Knowledge Discovery and Data
  Mining}.\hskip 1em plus 0.5em minus 0.4em\relax {ACM}, 2016, pp. 935--944.

\bibitem[Rosenbaum and Rubin(1983)]{RosenbaumR83}
P.~R. Rosenbaum and D.~B. Rubin, ``The central role of the propensity score in
  observational studies for causal effects,'' \emph{Biometrika}, vol.~70,
  no.~1, pp. 41--55, 1983.

\bibitem[Mikolov et~al.(2013)Mikolov, Sutskever, Chen, Corrado, and
  Dean]{word2vec}
T.~Mikolov, I.~Sutskever, K.~Chen, G.~Corrado, and J.~Dean, ``Distributed
  representations of words and phrases and their compositionality,'' in
  \emph{Annu. Conf. Neural Information Processing Systems (NIPS)}.\hskip 1em
  plus 0.5em minus 0.4em\relax Curran Associates Inc., 2013, p. 3111–3119.

\bibitem[Le and Mikolov(2014)]{paragraph2vec}
\BIBentryALTinterwordspacing
Q.~V. Le and T.~Mikolov, ``Distributed representations of sentences and
  documents,'' in \emph{Proc. 31st Int. Conf. Machine Learning (ICML)}, ser.
  {JMLR} Workshop and Conference Proceedings, vol.~32.\hskip 1em plus 0.5em
  minus 0.4em\relax JMLR.org, 2014, pp. 1188--1196. [Online]. Available:
  \url{http://proceedings.mlr.press/v32/le14.html}
\BIBentrySTDinterwordspacing

\bibitem[Devlin et~al.(2019)Devlin, Chang, Lee, and Toutanova]{devlin-bert}
J.~Devlin, M.~Chang, K.~Lee, and K.~Toutanova, ``{BERT:} pre-training of deep
  bidirectional transformers for language understanding,'' in \emph{Proc. Conf.
  North American Chapter of the Association for Computational Linguistics:
  Human Language Technologies (NAACL-HLT)}.\hskip 1em plus 0.5em minus
  0.4em\relax Association for Computational Linguistics, 2019, pp. 4171--4186.

\bibitem[Xu et~al.(2017)Xu, Liu, Feng, Yin, Song, and Song]{gemini}
X.~Xu, C.~Liu, Q.~Feng, H.~Yin, L.~Song, and D.~Song, ``Neural network-based
  graph embedding for cross-platform binary code similarity detection,'' in
  \emph{Proc. 2017 ACM SIGSAC Conf. Computer and Communications Security
  (CCS)}.\hskip 1em plus 0.5em minus 0.4em\relax ACM, 2017, p. 363–376.

\bibitem[Kim et~al.(2022)Kim, Kim, Cha, Son, and Kim]{bin-sim-lessons-learned}
D.~Kim, E.~Kim, S.~K. Cha, S.~Son, and Y.~Kim, ``Revisiting binary code
  similarity analysis using interpretable feature engineering and lessons
  learned,'' \emph{IEEE Trans. Software Eng.}, 2022.

\bibitem[Cai and Wang(2019)]{LocalDegreeProfile-graph-embedding}
C.~Cai and Y.~Wang, ``A simple yet effective baseline for non-attribute graph
  classification,'' in \emph{{ICLR} '19: International Conference on Learning
  Representations}, 2019.

\bibitem[Li et~al.(2019)Li, Wu, Guo, Liu, and Liu]{BoostNE}
J.~Li, L.~Wu, R.~Guo, C.~Liu, and H.~Liu, ``Multi-level network embedding with
  boosted low-rank matrix approximation,'' in \emph{{ASONAM} '19: Iternational
  Conference on Advances in Social Networks Analysis and Mining}.\hskip 1em
  plus 0.5em minus 0.4em\relax {ACM}, 2019, pp. 49--56.

\bibitem[Broder(1997)]{minhash}
A.~Z. Broder, ``On the resemblance and containment of documents,'' in
  \emph{Proc. Compression and Complexity of {SEQUENCES} 1997}.\hskip 1em plus
  0.5em minus 0.4em\relax {IEEE}, 1997, pp. 21--29.

\bibitem[Abadi et~al.(2015)Abadi, Agarwal, Barham, Brevdo, Chen, Citro,
  Corrado, Davis, Dean, Devin, Ghemawat, Goodfellow, Harp, Irving, Isard, Jia,
  Jozefowicz, Kaiser, Kudlur, Levenberg, Man\'{e}, Monga, Moore, Murray, Olah,
  Schuster, Shlens, Steiner, Sutskever, Talwar, Tucker, Vanhoucke, Vasudevan,
  Vi\'{e}gas, Vinyals, Warden, Wattenberg, Wicke, Yu, and
  Zheng]{tensorflow2015-whitepaper}
\BIBentryALTinterwordspacing
M.~Abadi, A.~Agarwal, P.~Barham, E.~Brevdo, Z.~Chen, C.~Citro, G.~S. Corrado,
  A.~Davis, J.~Dean, M.~Devin, S.~Ghemawat, I.~Goodfellow, A.~Harp, G.~Irving,
  M.~Isard, Y.~Jia, R.~Jozefowicz, L.~Kaiser, M.~Kudlur, J.~Levenberg,
  D.~Man\'{e}, R.~Monga, S.~Moore, D.~Murray, C.~Olah, M.~Schuster, J.~Shlens,
  B.~Steiner, I.~Sutskever, K.~Talwar, P.~Tucker, V.~Vanhoucke, V.~Vasudevan,
  F.~Vi\'{e}gas, O.~Vinyals, P.~Warden, M.~Wattenberg, M.~Wicke, Y.~Yu, and
  X.~Zheng, ``{TensorFlow}: Large-scale machine learning on heterogeneous
  systems,'' 2015, software available from tensorflow.org. [Online]. Available:
  \url{https://www.tensorflow.org/}
\BIBentrySTDinterwordspacing

\bibitem[Kneser and Ney(1995)]{kneser-ney}
R.~Kneser and H.~Ney, ``Improved backing-off for m-gram language modeling,'' in
  \emph{Int. Conf. Acoustics, Speech, and Signal Processing, (ICASSP)}.\hskip
  1em plus 0.5em minus 0.4em\relax {IEEE} Computer Society, 1995, pp. 181--184.

\bibitem[Chen and Goodman(1996)]{modified-kneser-ney}
S.~F. Chen and J.~Goodman, ``An empirical study of smoothing techniques for
  language modeling,'' in \emph{Proc. Annu. Meeting Association for
  Computational Linguistics (ACL)}.\hskip 1em plus 0.5em minus 0.4em\relax
  Morgan Kaufmann Publishers / {ACL}, 1996, pp. 310--318.

\bibitem[Heafield(2011)]{kenlm-queries}
K.~Heafield, ``{KenLM}: Faster and smaller language model queries,'' in
  \emph{Proc. Workshop on Statistical Machine Translation (WMT@EMNLP)}.\hskip
  1em plus 0.5em minus 0.4em\relax Association for Computational Linguistics,
  2011, pp. 187--197.

\bibitem[Bao et~al.(2014)Bao, Burket, Woo, Turner, and Brumley]{byteweight}
T.~Bao, J.~Burket, M.~Woo, R.~Turner, and D.~Brumley, ``{BYTEWEIGHT:} learning
  to recognize functions in binary code,'' in \emph{Proc. 23rd {USENIX}
  Security Symposium (USENIX Security)}.\hskip 1em plus 0.5em minus 0.4em\relax
  {USENIX} Association, 2014, pp. 845--860.

\bibitem[Shin et~al.(2015)Shin, Song, and Moazzezi]{shin2015}
E.~C.~R. Shin, D.~Song, and R.~Moazzezi, ``Recognizing functions in binaries
  with neural networks,'' in \emph{24th {USENIX} Security Symposium ({USENIX}
  Security 15)}.\hskip 1em plus 0.5em minus 0.4em\relax Washington, D.C.:
  {USENIX} Association, 2015, pp. 611--626.

\bibitem[Andriesse et~al.(2017)Andriesse, Slowinska, and Bos]{nucleus}
D.~Andriesse, A.~Slowinska, and H.~Bos, ``Compiler-agnostic function detection
  in binaries,'' in \emph{{IEEE} European Symposium on Security and Privacy
  (EuroS{\&}P)}.\hskip 1em plus 0.5em minus 0.4em\relax {IEEE}, 2017, pp.
  177--189.

\bibitem[David et~al.(2020)David, Alon, and Yahav]{nero}
Y.~David, U.~Alon, and E.~Yahav, ``Neural reverse engineering of stripped
  binaries using augmented control flow graphs,'' \emph{Proc. {ACM} Program.
  Lang.}, vol.~4, no. {OOPSLA}, pp. 225:1--225:28, 2020.

\bibitem[Lacomis et~al.(2019)Lacomis, Yin, Schwartz, Allamanis, Goues, Neubig,
  and Vasilescu]{dire}
J.~Lacomis, P.~Yin, E.~J. Schwartz, M.~Allamanis, C.~L. Goues, G.~Neubig, and
  B.~Vasilescu, ``{DIRE:} {A} neural approach to decompiled identifier
  naming,'' in \emph{34th {IEEE/ACM} International Conference on Automated
  Software Engineering (ASE)}.\hskip 1em plus 0.5em minus 0.4em\relax {IEEE},
  2019, pp. 628--639.

\bibitem[Yang et~al.(2021)Yang, Guo, Hao, Ciptadi, Ahmadzadeh, Xing, and
  Wang]{CADE}
L.~Yang, W.~Guo, Q.~Hao, A.~Ciptadi, A.~Ahmadzadeh, X.~Xing, and G.~Wang,
  ``{CADE:} detecting and explaining concept drift samples for security
  applications,'' in \emph{30th {USENIX} Security Symposium, {USENIX} Security
  2021, August 11-13, 2021}.\hskip 1em plus 0.5em minus 0.4em\relax {USENIX}
  Association, 2021, pp. 2327--2344.

\bibitem[Schrittwieser et~al.(2016)Schrittwieser, Katzenbeisser, Kinder,
  Merzdovnik, and Weippl]{obfuscation-csur16}
S.~Schrittwieser, S.~Katzenbeisser, J.~Kinder, G.~Merzdovnik, and E.~Weippl,
  ``Protecting software through obfuscation: Can it keep pace with progress in
  code analysis?'' \emph{ACM Computing Surveys}, vol.~49, no.~1, 2016.

\bibitem[Hu et~al.(2021)Hu, Ke, Wang, and Lyu]{HuKWL21}
S.~Hu, L.~Ke, X.~Wang, and S.~Lyu, ``Tkml-ap: Adversarial attacks to top-k
  multi-label learning,'' in \emph{Proc. IEEE/CVF Int. Conf. on Computer Vision
  (ICCV)}, 2021, pp. 7649--7657.

\bibitem[Song et~al.(2018)Song, Jin, Huang, and Hu]{SongJHH18}
Q.~Song, H.~Jin, X.~Huang, and X.~Hu, ``Multi-label adversarial
  perturbations,'' in \emph{{IEEE} Int. Conf. Data Mining (ICDM)}.\hskip 1em
  plus 0.5em minus 0.4em\relax {IEEE} Computer Society, 2018, pp. 1242--1247.

\bibitem[Castro et~al.(2021)Castro, Mu{\~{n}}oz{-}Gonz{\'{a}}lez, Pendlebury,
  Rodosek, Pierazzi, and Cavallaro]{CastroGPRPC21}
\BIBentryALTinterwordspacing
R.~L. Castro, L.~Mu{\~{n}}oz{-}Gonz{\'{a}}lez, F.~Pendlebury, G.~D. Rodosek,
  F.~Pierazzi, and L.~Cavallaro, ``Universal adversarial perturbations for
  malware,'' \emph{CoRR}, vol. abs/2102.06747, 2021. [Online]. Available:
  \url{https://arxiv.org/abs/2102.06747}
\BIBentrySTDinterwordspacing

\bibitem[Carlini et~al.(2019)Carlini, Athalye, Papernot, Brendel, Rauber,
  Tsipras, Goodfellow, Madry, and Kurakin]{Carlini-robustness19}
N.~Carlini, A.~Athalye, N.~Papernot, W.~Brendel, J.~Rauber, D.~Tsipras, I.~J.
  Goodfellow, A.~Madry, and A.~Kurakin, ``On evaluating adversarial
  robustness,'' \emph{CoRR}, vol. abs/1902.06705, 2019.

\bibitem[Duan et~al.(2020)Duan, Li, Wang, and Yin]{duan2020deepbindiff}
Y.~Duan, X.~Li, J.~Wang, and H.~Yin, ``Deepbindiff: Learning program-wide code
  representations for binary diffing,'' in \emph{Annu. Network and Distributed
  System Security Symp. (NDSS)}.\hskip 1em plus 0.5em minus 0.4em\relax The
  Internet Society, 2020.

\bibitem[Liu et~al.(2018)Liu, Huo, Zhang, Li, Li, Piao, and Zou]{alphadiff}
B.~Liu, W.~Huo, C.~Zhang, W.~Li, F.~Li, A.~Piao, and W.~Zou,
  ``{\(\alpha\)}diff: cross-version binary code similarity detection with
  {DNN},'' in \emph{Proc. 33rd {ACM/IEEE} Int. Conf. Automated Software
  Engineering (ASE)}.\hskip 1em plus 0.5em minus 0.4em\relax {ACM}, 2018, pp.
  667--678.

\bibitem[Zuo et~al.(2019)Zuo, Li, Young, Luo, Zeng, and Zhang]{zuo2019neural}
F.~Zuo, X.~Li, P.~Young, L.~Luo, Q.~Zeng, and Z.~Zhang, ``Neural machine
  translation inspired binary code similarity comparison beyond function
  pairs,'' in \emph{Annu. Network and Distributed System Security Symp.
  (NDSS)}.\hskip 1em plus 0.5em minus 0.4em\relax The Internet Society, 2019.

\bibitem[Sun et~al.(2019)Sun, Chen, Shan, Wang, et~al.]{sun2019binary}
W.~Sun, Y.~Chen, Z.~Shan, Q.~Wang \emph{et~al.}, ``Binary semantic similarity
  comparison based on software gene,'' in \emph{Journal of Physics: Conference
  Series}, vol. 1325, no.~1.\hskip 1em plus 0.5em minus 0.4em\relax IOP
  Publishing, 2019, p. 012109.

\bibitem[Bielik et~al.(2016)Bielik, Raychev, and Vechev]{phog}
P.~Bielik, V.~Raychev, and M.~T. Vechev, ``{PHOG:} probabilistic model for
  code,'' in \emph{Proc. 33nd Int. Conf. Machine Learning (ICML)}, ser. {JMLR}
  Workshop and Conference Proceedings, vol.~48.\hskip 1em plus 0.5em minus
  0.4em\relax JMLR.org, 2016, pp. 2933--2942.

\bibitem[Alon et~al.(2018)Alon, Zilberstein, Levy, and Yahav]{alon18}
U.~Alon, M.~Zilberstein, O.~Levy, and E.~Yahav, ``A general path-based
  representation for predicting program properties,'' in \emph{Proc. 39th {ACM}
  {SIGPLAN} Conf. Programming Language Design and Implementation (PLDI)}.\hskip
  1em plus 0.5em minus 0.4em\relax {ACM}, 2018, pp. 404--419.

\bibitem[Brockschmidt et~al.(2019)Brockschmidt, Allamanis, Gaunt, and
  Polozov]{brockschmidt2018generative}
\BIBentryALTinterwordspacing
M.~Brockschmidt, M.~Allamanis, A.~L. Gaunt, and O.~Polozov, ``Generative code
  modeling with graphs,'' in \emph{7th Int. Conf. Learning Representations
  (ICLR)}.\hskip 1em plus 0.5em minus 0.4em\relax OpenReview.net, 2019.
  [Online]. Available: \url{https://openreview.net/forum?id=Bke4KsA5FX}
\BIBentrySTDinterwordspacing

\bibitem[Allamanis et~al.(2018)Allamanis, Brockschmidt, and
  Khademi]{allamanis2018learning}
M.~Allamanis, M.~Brockschmidt, and M.~Khademi, ``Learning to represent programs
  with graphs,'' in \emph{International Conference on Learning
  Representations}, 2018.

\bibitem[Fernandes et~al.(2019)Fernandes, Allamanis, and
  Brockschmidt]{fernandes2018structured}
\BIBentryALTinterwordspacing
P.~Fernandes, M.~Allamanis, and M.~Brockschmidt, ``Structured neural
  summarization,'' in \emph{7th Int. Conf. Learning Representations
  (ICLR)}.\hskip 1em plus 0.5em minus 0.4em\relax OpenReview.net, 2019.
  [Online]. Available: \url{https://openreview.net/forum?id=H1ersoRqtm}
\BIBentrySTDinterwordspacing

\bibitem[Alon et~al.(2019)Alon, Brody, Levy, and Yahav]{alon2018codeseq}
\BIBentryALTinterwordspacing
U.~Alon, S.~Brody, O.~Levy, and E.~Yahav, ``code2seq: Generating sequences from
  structured representations of code,'' in \emph{7th Int. Conf. Learning
  Representations (ICLR)}.\hskip 1em plus 0.5em minus 0.4em\relax
  OpenReview.net, 2019. [Online]. Available:
  \url{https://openreview.net/forum?id=H1gKYo09tX}
\BIBentrySTDinterwordspacing

\bibitem[Allamanis et~al.(2016)Allamanis, Peng, and
  Sutton]{allamanis2016convolutional}
M.~Allamanis, H.~Peng, and C.~Sutton, ``A convolutional attention network for
  extreme summarization of source code,'' in \emph{Proc. 33nd Int. Conf.
  Machine Learning (ICML)}, ser. {JMLR} Workshop and Conference Proceedings,
  vol.~48.\hskip 1em plus 0.5em minus 0.4em\relax JMLR.org, 2016, pp.
  2091--2100.

\bibitem[Kalgutkar et~al.(2019)Kalgutkar, Kaur, Gonzalez, Stakhanova, and
  Matyukhina]{kalgutkar2019code}
V.~Kalgutkar, R.~Kaur, H.~Gonzalez, N.~Stakhanova, and A.~Matyukhina, ``Code
  authorship attribution: Methods and challenges,'' \emph{ACM Computing Surveys
  (CSUR)}, vol.~52, no.~1, pp. 1--36, 2019.

\end{thebibliography}

\end{document}